\documentclass[aip,pof,reprint,onecolumn]{revtex4-1} %twocolumn
\usepackage{graphicx}% Include figure files
\usepackage{dcolumn}% Align table columns on decimal point
\usepackage{bm}% bold math
\usepackage{amssymb}
\usepackage{amsmath}
\usepackage{amsfonts}
\usepackage{mathtools}
\usepackage{comment}
\usepackage{lineno}
\usepackage{soul}
\usepackage{natbib}
\usepackage{color}     
\usepackage{soul}
\usepackage[usenames,dvipsnames,svgnames,table]{xcolor} 
\usepackage[colorlinks=true,
            linkcolor=blue,
            urlcolor=blue,
            citecolor=BrickRed]{hyperref}
            
\usepackage{mathrsfs}  

\usepackage{tabulary}

\newcommand{\black}{\color{black}}

\usepackage{booktabs}

\usepackage{mathtools}
\usepackage{amssymb}
\usepackage{amsmath}

\usepackage[abs]{overpic}
\usepackage{xcolor,varwidth}   
\definecolor{light-gray}{gray}{0.95}
\newcommand{\code}[1]{\colorbox{light-gray}{\texttt{#1}}}

\begin{document}
\title{Data-driven quantification of model-form uncertainty in Reynolds-averaged simulations of wind farms}
%\author{...}
\author{Ali Eidi}
\affiliation{School of Civil Engineering, College of Engineering, University of Tehran, Tehran, Iran}
\author{Navid Zehtabiyan-Rezaie}
\affiliation{Department of Mechanical and Production Engineering, Aarhus University, 8000 Aarhus C, Denmark}
\author{Reza Ghiassi}
\email{rghiassi@ut.ac.ir}
\affiliation{School of Civil Engineering, College of Engineering, University of Tehran, Tehran, Iran}
\author{Xiang Yang}
\affiliation{Department of Mechanical Engineering, Pennsylvania State University, State College, PA, 16802, USA}
\author{Mahdi Abkar}
\email{abkar@mpe.au.dk}
\affiliation{Department of Mechanical and Production Engineering, Aarhus University, 8000 Aarhus C, Denmark}

\begin{abstract}
Computational fluid dynamics using the Reynolds-averaged Navier–Stokes (RANS) remains the most cost-effective approach to study wake flows and power losses in wind farms. The underlying assumptions associated with turbulence closures are one of the biggest sources of errors and uncertainties in the model predictions. This work aims to quantify model-form uncertainties in RANS simulations of wind farms at high Reynolds numbers under neutrally stratified conditions by perturbing the Reynolds stress tensor through a data-driven machine-learning technique. 
To this end, a two-step feature-selection method is applied to determine key features of the model. Then, the extreme gradient boosting algorithm is validated and employed to predict the perturbation amount and direction of the modeled Reynolds stress toward the limiting states of turbulence on the barycentric map. This procedure leads to a more accurate representation of the Reynolds stress anisotropy.
The data-driven model is trained on high-fidelity data obtained from large-eddy simulation of a specific wind farm, and it is tested on two other (unseen) wind farms with distinct layouts to analyze its performance in cases with different turbine spacing and partial wake. 
The results indicate that, unlike the data-free approach in which a uniform and constant perturbation amount is applied to the entire computational domain, the proposed framework yields an optimal estimation of the uncertainty bounds for the RANS-predicted quantities of interest, including the wake velocity, turbulence intensity, and power losses in wind farms.   
\end{abstract}

\maketitle

%%% main text
\section{Introduction} \label{sec:Introduction}
The world's demand for electricity rises annually, and in the same direction, the installed capacity of the cleaner power production methods increase to mitigate global warming \cite{papadis2020challenges}. Wind energy - one of the most promising clean power production methods - can play a vital role in the transition path toward net-zero greenhouse emissions, but there exist several challenges. These challenges are primarily associated with the complex flow in wind farms due to the two-way interactions between the atmospheric boundary layer (ABL) and the turbines. 
The structures in a wind-farm flow vary from turbine scale to meteorological scale \cite{stevens2017flow,veers2019grand,PorteAgel2019Review}. Studying the wind-farm flow is one of the necessary steps in finding the optimal design and operation of wind farms.
However, due to the enormous complexity of wind-farm flow, the investigations using high-fidelity computational fluid dynamics (CFD) techniques, e.g., large-eddy simulation (LES), are very computationally expensive.  
Therefore, mid-fidelity methods, including Reynolds-averaged Navier–Stokes (RANS) models, are the most widely used CFD tools in the wind-energy industry \cite{vermeer2003wind}.
Despite their lower computational costs compared to LESs, the RANS models introduce considerable errors and uncertainties in measuring the quantities of interest (QoIs) in realistic flow modelings due to the inherent assumptions in their structure.  
A part of the inconsistencies in predictions of the RANS models originates from the linear eddy-viscosity hypothesis, which leads to discrepancies in the calculated Reynolds stresses  \cite{xiao2019quantification}. This can be partly addressed by applying eigenvalue perturbation to the Reynolds stress anisotropy to quantify the model-form uncertainties \cite{emory2013modeling}.

\textit{A posteriori} analysis for data-free uncertainty quantification (UQ) via the eigenvalue perturbation technique has been followed in recent studies, e.g., 
urban-flow modeling \cite{gorle2015quantifying}, flow over backward-facing step \cite{cremades2019reynolds}, pollutant dispersion \cite{garcia2017quantifying}, and flow on a wavy wall \cite{gorle2019epistemic}, among others.
Turning to wind-energy-related studies, Hornsh{\o}j-M{\o}ller et al. \cite{hornshoj2021quantifying} applied this technique for a case with a standalone turbine subjected to the ABL flow on flat terrain. They investigated different amounts of perturbation values and assessed the normalized velocity deficit and turbulence intensity in the wake region. Eidi et al. \cite{eidi2021model} attempted to quantify model-from uncertainty in the RANS simulation of wind farms. They applied a similar method to a group of consecutive turbines in a wind farm using constant perturbation values toward different limiting states of turbulence for the entire domain. Two other wind-farm layouts were studied, and the framework was shown to provide adequate coverage over the results of power losses predicted by the high-fidelity model.
However, since the optimal amount of perturbation is not the same for all computational cells, this strategy was unable to reduce uncertainty while retaining high accuracy.

Nowadays, the data-driven models along with machine learning (ML) techniques are being extensively utilized in fluid-mechanics-related problems. Practical methods for creating a mapping between large data-sets and QoIs are provided by ML which can be used as a standalone tool for supervised or unsupervised learning, or it can be combined with other models (e.g., statistical inference), to offer $a$ $posteriori$ corrections \cite{duraisamy2019turbulence,Brunton2020,tracey2013application,duraisamy2015new,singh2017machine}.
Although the interpretability and generalizability of data-driven models are among the most challenging issues, these models can reduce the need for computationally expensive physics-based models \cite{legaard2021,Brunton2022,vinuesa2022enhancing}.
Several studies are using ML techniques, specifically deep learning tools, for predicting fluid-mechanics-related problems by utilizing mid- and high-fidelity data from RANS simulations and LESs (or direct numerical simulations), e.g., flow over periodic hill \cite{xiao2020flows, schmelzer2020discovery,li2022data}, wall mounted cube \cite{ling2016machine}, backward-facing step \cite{kaandorp2020data}, duct flow \cite{wu2018physics, wang2017comprehensive, ling2015evaluation,jiang2021interpretable}, RANS correction \cite{steiner2020data, luan2020influence}, flow with system rotation \cite{huang2021bayesian}, and wind-farm flow modeling \cite{ti2020wake, ali2021cluster, ali2021clustering}, among others. For more information, see, e.g., reviews of Refs. \cite{duraisamy2021perspectives, durbin2018some, zehtab2022, ahmed2021closures} and references therein.
A few examples of the ML-assisted studies include direct prediction of the Reynolds stress \cite{geneva2019quantifying}, and estimation of the Reynolds stress and eddy-viscosity discrepancies between low and high-fidelity approaches \cite{ling2015evaluation}. Turning to the UQ analysis, Heyse et al. \cite{heyse2021estimating} quantified uncertainty in RANS modeling of the flow over a wavy wall via the eigenvalue perturbation by utilizing the random forest (RF) algorithm. They used $a$ $priori$ determined local perturbation value based on the distance between the RANS and LES results on a barycentric map, yet perturbing the eigenvalues toward all turbulence limiting states.

To sum up, in the studies focusing on wind-turbine and wind-farm wake, UQ based on the eigenvalue perturbation has been performed by utilizing a data-free approach to compute the perturbation value for the whole domain.
With the growing number of available high-fidelity data, the data-driven UQ in wind-farm flow modeling could be of interest to the wind-energy community. The data-driven approach could estimate the optimal amount of perturbation in each direction.
To the best knowledge of the authors, ML-assisted UQ in wind-farm flow modeling has not been addressed yet.
To bridge the research gap, this study aims to investigate a novel data-driven framework associated with the amount and direction of eigenvalue perturbation of the Reynolds stress anisotropy tensor. This approach utilizes an ML algorithm to surrogate a precise method for characterizing uncertainty in the RANS simulations of wind farms. It also attempts to alleviate the generalizability issue in ML-predicted results by utilizing a two-step feature-selection procedure and a robust model.
The rest of the paper is laid out as follows. 
In \autoref{sec:method}, the methodology of this study is presented.
In \autoref{sec:R&D}, data-driven results for the labeled data are shown and compared with data-free predictions, then the feature-selection procedure is explained. The rest of this section deliberates the procedure of validating the ML model and its predictions. Finally, essential conclusions are highlighted in \autoref{sec:conc}.

\section{Methodology} \label{sec:method}
In this section, we first discuss the eigenvalue perturbation method and then attempt to improve it by utilizing a data-driven approach. Then, the CFD models used in this study are thoroughly discussed, and the ML algorithm is introduced with a detailed analysis of the input features and output targets.

\subsection{Mathematical details of uncertainty quantification} \label{sec:UQ md}

\begin{figure}[ht]
    \centering
    \includegraphics[width=0.6\linewidth]{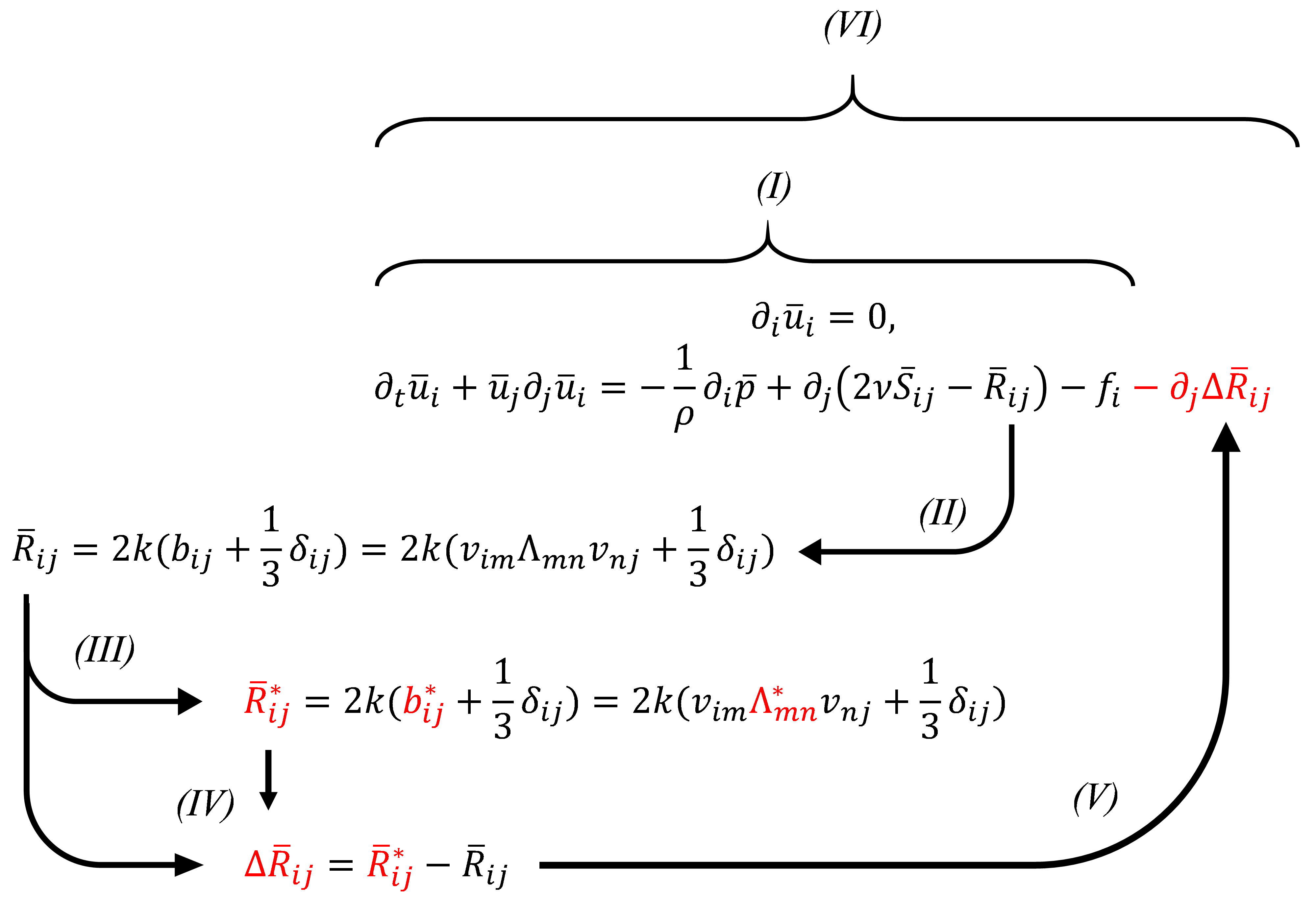}
    \caption{An overview of the RANS perturbation procedure.}
    \label{fig:RANS_UQ}
\end{figure}

For an incompressible turbulent flow, the Reynolds-averaged continuity and momentum equations are derived as $(I)$ in \autoref{fig:RANS_UQ}. 
The mean pressure, mean velocity, density, kinematic viscosity, and the turbine-induced force are represented by $\bar{p}$, $\bar{u}$, $\rho$, $\nu$, and $f_i$, respectively. Here, $\Bar{R}_{ij}$ refers to the Reynolds stress quantified by the Boussinesq's hypothesis as $2/3 k \delta_{ij}-2\nu_{T}\bar{S}_{ij}$, where $k$= 0.5 trace($\Bar{R}_{ij}$), $\delta_{ij}$, and $\nu_{T}$ are the turbulent kinetic energy, the Kronecker delta, and turbulent viscosity, respectively, and $\bar{S}_{ij}$ denotes the mean strain rate tensor.
In the two-equation models, the Reynolds stress term is derived using two additional transport equations \cite{oztop2012analysis}. Similar to our previous study \cite{eidi2021model}, we utilize the realizable $k-\varepsilon$ model \cite{shih1995new} for the RANS simulation of wind farms on flat terrain.

The Reynolds averaging, the Reynolds stress representation, functional forms, and model coefficients can lead to uncertainty in the predictions of the RANS models \cite{duraisamy2019turbulence}; some of which are intractable, but in some cases, the uncertainty can be reduced or even quantified. 
In this research, we utilize Reynolds stress anisotropy to investigate the model-form uncertainty in RANS
simulation of wind farms. It is worth highlighting that Reynolds stress anisotropy in wind-turbines wake has been studied in the past (see, e.g. Refs. \cite{Naseem2018, ali2019classification, camp2019low}).
By eigen-decomposition of the normalized Reynolds stress anisotropy \cite{kreyszig2009advanced}, one can rewrite it as the product of its eigenvalues ($\lambda$) and eigenvectors ($v$) representing the shape and orientation of the tensor, respectively  ($(II)$ in \autoref{fig:RANS_UQ}) \cite{emory2013modeling}, where $\Lambda_{mn}$ is a diagonal tensor with $\lambda_i$ as its components, and $v_{ij}$ is the eigenvector tensor \cite{banerjee2007presentation}. The $\lambda_i$ are sorted in descending order and satisfying the normalization constraints, i.e.,
% $\lambda_{1}+\lambda_{2}+\lambda_{3}=0$.
trace($\Lambda_{mn}$)=0. 

 \begin{figure}[ht]
    \centering
    \includegraphics[width=0.45\linewidth]{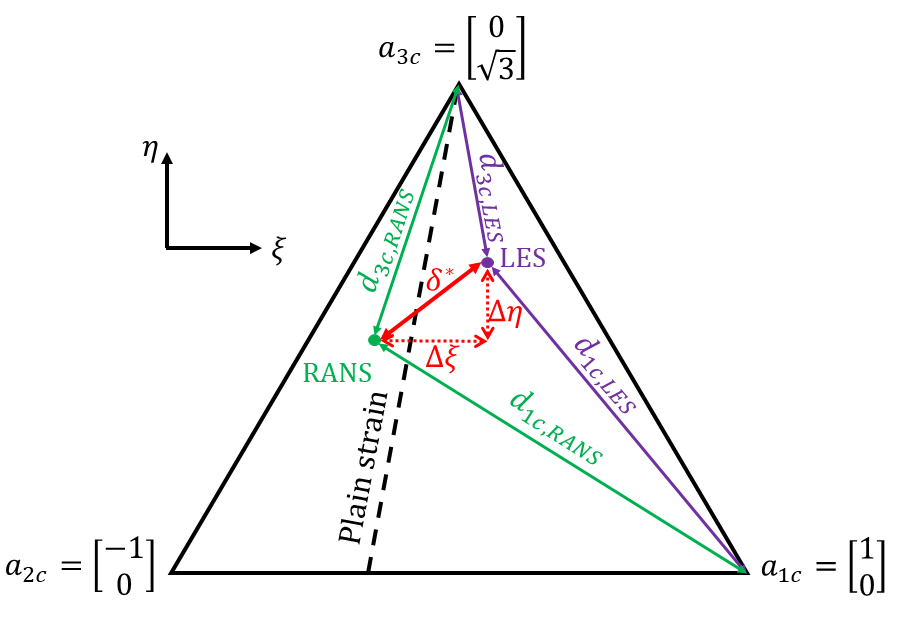}
    \caption{Barycentric triangle with the corresponding points of LES and RANS.}
    \label{fig:Bary_tri}
\end{figure}

As shown in \autoref{fig:Bary_tri}, the Reynolds stress anisotropy can be mapped on a barycentric coordinate represented by an equilateral triangle to be illustrated in a two-dimensional space \cite{Naseem2018}.
Each corner of the triangle represents the limiting states of one- ($a_{1c}$), two- ($a_{2c}$), and three-component turbulence ($a_{3c}$).
One can have a perturbed Reynolds stress with a varied shape by considering the location of the RANS Reynolds stress anisotropy on this triangle. Perturbing with a specific perturbation value ($\delta$) toward each corner using limiting state eigenvalue tensor (i.e., $\mathbf{\Lambda}_{c}$ \cite{hornshoj2021quantifying})
leads to a perturbed tensor of the eigenvalues ($\Lambda^*$) as 

\begin{equation}
    \begin{gathered}
        \label{eq:delta_perturb}
            \Lambda^* = \Lambda + \delta(\Lambda_{c}-\Lambda).
    \end{gathered}
\end{equation}

The perturbed Reynolds stress ($\Bar{R^*}_{ij}$) is derived when $\Lambda^*$ is assigned to the anisotropy tensor equations given as $(III)$ in \autoref{fig:RANS_UQ}.
Then, including the difference between the baseline and the perturbed Reynolds stresses in the momentum equations as a divergence source term results in a perturbed model as given in $(IV)$ and $(V)$ in \autoref{fig:RANS_UQ}. Finally, solving the perturbed RANS equations ($(VI)$ in \autoref{fig:RANS_UQ}) predicts the perturbed QoIs.
We performed this approach in the prior study \cite{eidi2021model} by applying a constant perturbation value ($\delta$) to all computational cells in the domain and toward all three limiting states.
The results of the perturbed model covered the LES outputs within a bound for all QoIs, where one- and three-component perturbations were the upper or lower limits, and two-component results always fell inside the band. However, in this study, we are pursuing three specific objectives regarding the eigenvalue perturbation of the Reynolds stress anisotropy.
First, we intend to determine the perturbation direction for individual cells of the domain. As the bandwidth of the perturbed results quantifies the uncertainty level, we aim for an optimal bound to determine the QoIs with higher accuracy (i.e., decreasing the uncertainty bandwidth and covering the high-fidelity results simultaneously). Therefore, a perturbation toward $a_{2c}$ - a theoretically limiting state between one and three-component turbulence \cite{iaccarino2017eigenspace}- does not affect the UQ bandwidth and is no longer required. By having the location of the LES and RANS Reynolds stress anisotropy, one can determine the perturbation direction. In other words, by comparing the distance between the RANS and LES points with each of $a_{1c}$ and $a_{3c}$ on the barycentric map (i.e., $d_{1c, RANS}$ vs. $d_{1c, LES}$ and $d_{3c, RANS}$ vs. $d_{3c, LES}$ in \autoref{fig:Bary_tri}), the RANS point could be perturbed toward a corner to which the LES point is closer than the RANS. This will be further discussed in \autoref{sec:predict}. 

The second objective of the current study is to assign exclusive $\delta$ values for the cells. One can consider two methods of data-free and data-driven, to achieve this objective. In the data-free approach, a constant $\delta$ is considered for the whole domain determined by a trial and error procedure, while in the second approach, data-driven methods are used to estimate appropriate $\delta$ for each domain cell. The current study seeks to determine $\delta$ for individual cells. To this end, we estimate the distance between the corresponding points of LES and RANS on the barycentric map ($\delta^*$) considering realizability conditions (i.e., $\delta = \min [\delta^*, 1]$). Briefly, each cell in the domain is perturbed toward either of $a_{1c}$ or $a_{3c}$ using an exclusively determined $\delta$. 

The third objective is to selectively apply the perturbation to the cells. Achieving this objective can play an essential role in minimizing the computational cost. For the free stream zone of the domain, the calculation yields a $\delta$ value of approximately zero, which corresponds to no perturbation. The normalized velocity deficit, defined as $\Delta \bar{u}/\bar{u}_{h}$, could be utilized as a criterion to determine whether cells are inside or outside the free stream zone, where $\Delta \bar{u}=\bar{u}_{in}-\bar{u}$, and $\bar{u}_{in}$ and $\bar{u}_{h}$ are the inlet and hub height velocities, respectively. In other words, only the domain's required cells (i.e., wake zone) are perturbed with a given $\delta$ toward one of the limiting states. For further details regarding the cell zone detection, see \textcolor{blue}{Appendix} \ref{app:A}.

\subsection{Modeling cases and simulation details} \label{sec:cases}
\begin{figure} [ht]
    \centering
    \includegraphics[width=0.9\linewidth]{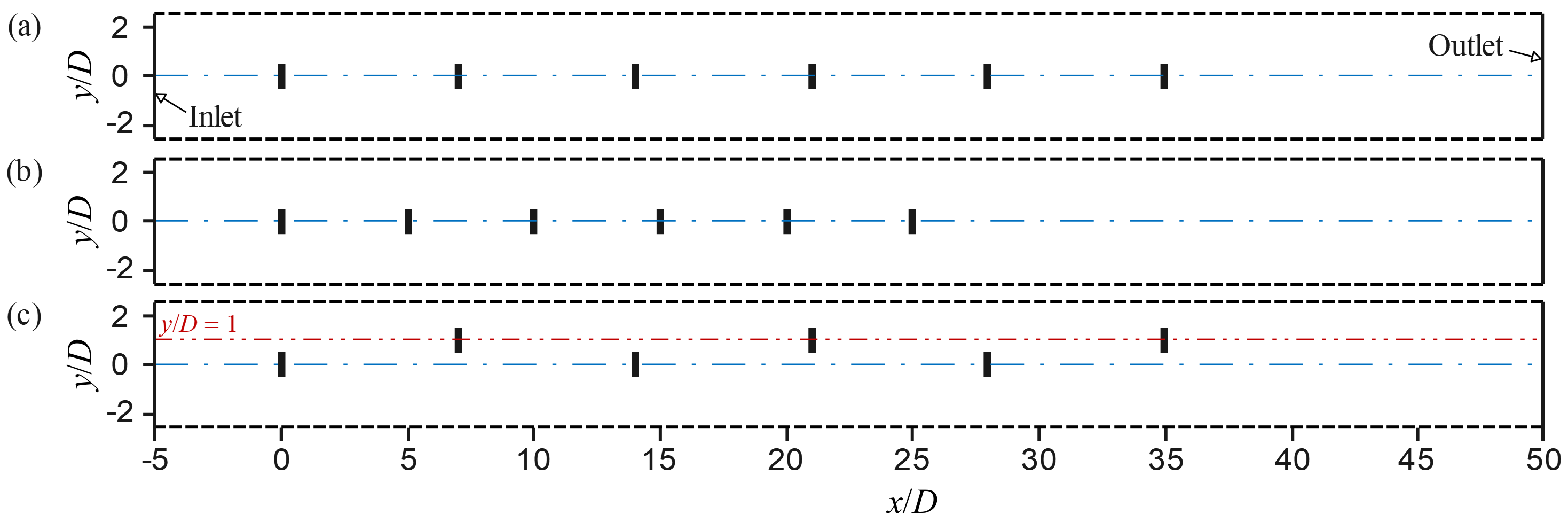}
    \caption{Three different layouts for the wind farm domain: (a) Case A, (b) Case B, and (c) Case C. Black dashed lines indicate cyclic boundaries, and the rectangles show the turbines' locations.}
    \label{fig:domains}
\end{figure}

This study focuses on three different wind-farm layouts (Cases A, B, and C), each including six rows of turbines with configurations shown in \autoref{fig:domains}.
The first turbine's distance from the domain inlet is set to be equal to $5D$ for all cases, where $D$ is the turbine's rotor diameter.
In Case A, the turbines are spaced $7D$ apart from their upstream counterparts.
The turbine spacing is equal to $5D$ in Case B. The second case is considered to investigate a condition with a closer turbine spacing resulting in increased turbulence and velocity deficit in the wake region. In order to study the performance of the proposed framework in a case with partial wake, in Case C, the even turbine rows are relocated $1D$ in the positive $y$-direction. In the $x$-direction, Case C has the same turbine spacing as that of Case A.

In the RANS framework, the computational domain has a dimension of 4400 m $\times$ 400 m $\times$ 355 m in the $x$, $y$, and $z$ directions, respectively. According to the grid-dependency analysis performed in our previous study \cite{eidi2021model}, a nonuniform mesh with 234 and 58 cells in the $x$ and $z$ directions and a uniform grid with $\Delta y$ = 10 m in the $y$-direction is used for the three cases. 
A non-rotating actuator-disk model \cite{Calaf2010} with an induction coefficient of $a=0.25$, and a modified thrust coefficient of $C^\prime_T =4a /(1-a)=$ 4/3 based on the velocity at the rotor area, is used to model the turbines with hub height of $z_{h}$ = 70 m and $D$ = 80 m. 
At the inlet boundary, the ABL logarithmic equation \cite{van2019improved} is used to compute the airflow velocity with $u_{h}$ = 8 m/s, and at the outlet, a zero-gradient condition is imposed.
The no-slip condition is satisfied on the ground (i.e., flat terrain), and the logarithmic velocity profile is used to adjust the velocity at the top boundary.
The zero-gradient condition is used for the pressure at the inlet, top, and ground boundaries, while the relative pressure at the outlet is set to zero.
The incoming turbulence intensity at the hub height ($I=\sqrt{2k/3}/\bar{u}_{h}$) is 5.8\%, and the value of the aerodynamic surface roughness ($z_{0}$) is adjusted to be consistent with these quantities.
Appropriate wall functions for $k$, $\varepsilon$, and $\nu_{T}$ are utilized to predict the optimal values regarding the ABL velocity profile on the ground boundary \cite{hargreaves2007use}.
For all flow parameters, the cyclic boundary condition is applied at the side walls.
The governing equations for the baseline and perturbed RANS models (($I$) and ($VI$) in \autoref{fig:RANS_UQ}) are solved using the finite volume method
implemented in a solver in the open-source package of OpenFOAM v2006. To handle the pressure-velocity coupling, the steady-state semi-implicit pressure linked equation (SIMPLE) algorithm \cite{ferziger2002computational} is used. 

The computational domain in the LES framework is uniformly divided into 480 $\times$ 160 $\times$ 72 grid points and the spatial resolution is 10 m, 5 m, and 5 m in the streamwise ($x$), spanwise ($y$), and wall-normal ($z$) directions. 
For the streamwise and spanwise directions, a pseudo-spectral scheme is utilized to calculate the spatial derivatives, and in the wall-normal direction, a second-order finite-difference technique is used. 
A second-order-accurate Adams-Bashforth scheme \cite{Canuto1988} is employed for the integration in time.
In the wall-normal direction, the grid planes are staggered, with the first vertical velocity plane located on the surface and the first horizontal velocity plane 2.5 m away from it. 
In the horizontal directions, the boundary conditions are cyclical.
We have utilized a fringe zone to adapt the downstream wake flow with the undisturbed flow condition at the inlet \cite{stevens2014concurrent}. Hence, the LES domain is marginally extended (+ 400 m) in the streamwise direction compared to the domain used for the RANS simulations.
The instantiated surface stress for the ground is computed employing the local application of the Monin-Obukhov similarity theory \cite{Moeng1984,Yang2018}.
A zero vertical-velocity along with a stress-free condition is assigned to the top boundary.
This LES code has been well validated and used in prior studies. The reader may refer to Refs. \cite{Porte-Agel2011,Abkar2013,Wu2015,Abkar2015a,bastankhah2019multirotor} for further details of the LES framework and the solver.

\subsection{XGBoost algorithm} \label{sec:XGBoost}

 \begin{figure}%[ht]
    \centering
    \includegraphics[width=0.85\linewidth]{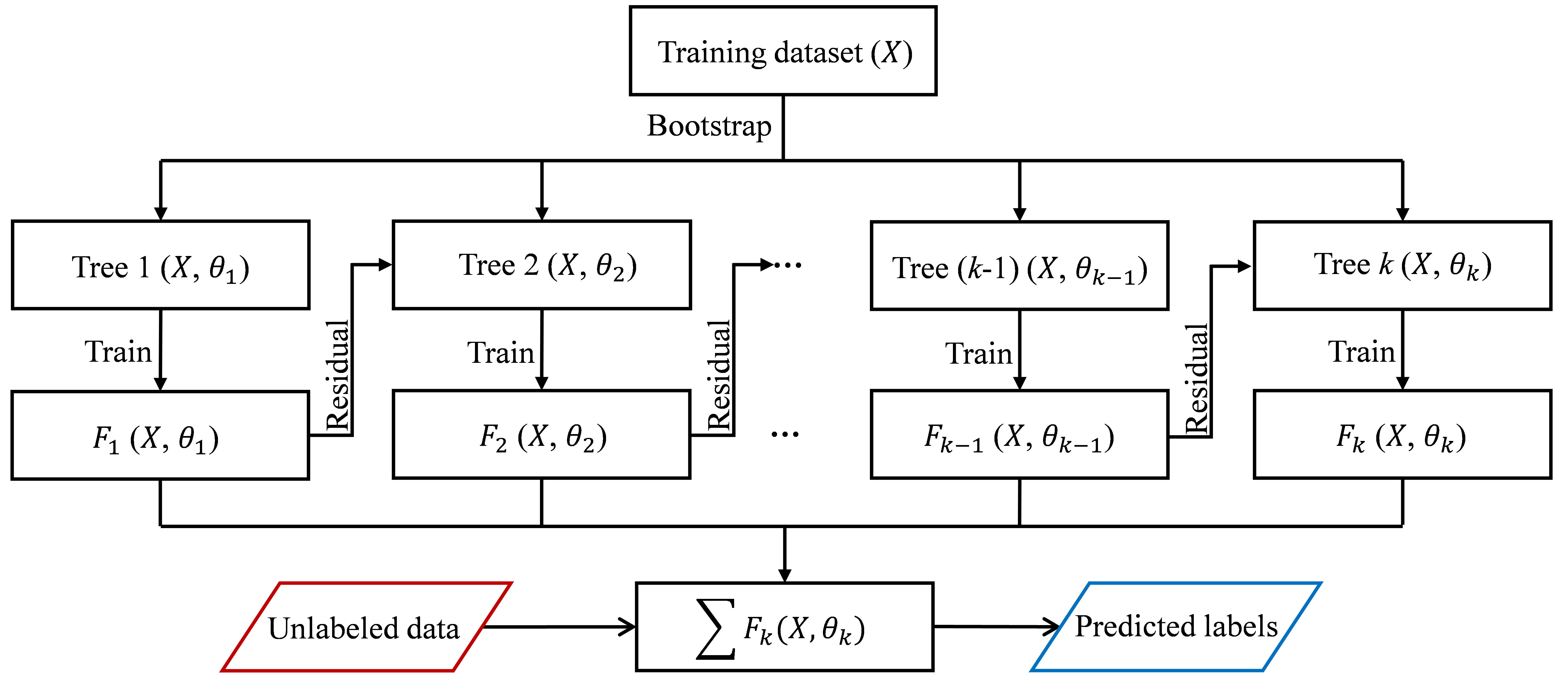}
    \caption{The flowchart of XGBoost \cite{guo2020degradation} (\textit{redrawn}).}
    \label{fig:xgb}
\end{figure}

The extreme gradient boosting (XGBoost) is an efficient and scalable application of the well-known gradient-boosted decision-tree technique developed by Chen and Guestrin \cite{chen2016xgboost}. It combines decision-tree classifiers with regressions to achieve the best results. As in the preceding scenario, numerous trees are utilized to learn separate output predictions, which are then aggregated.
Through an iterative procedure, the gradient boosting enhances poor base learning models \cite{friedman2001greedy}.
The residual can be used to optimize a loss function, as illustrated in \autoref{fig:xgb}.
In XGBoost, the objective function for monitoring the performance of the model is

\begin{equation}
    \begin{gathered}
        \label{eq:XG1}
            Obj(\Theta)= F_{loss}(\Theta) + F_{\lambda}(\Theta), 
    \end{gathered}
\end{equation}
where $\Theta$, $Obj(\Theta)$, $F_{loss}(\Theta)$, and $F_{\lambda}(\Theta)$ are the parameter learned from data, the objective function, the training loss function (e.g., square loss), and a regularization term (to avoid over-fitting, e.g., $L1$ or $L2$ norm), respectively \cite{zhang2018data}. 
The output of the model is ${\hat y}_{i}= \sum_{i=1}^{k} f_{k}(x_{i})$, in which $f_{k} \in F$ is averaged or voted by a collection $F$ of $k$ trees.
Then, the objective function for the $t^{th}$ iteration can be represented as

\begin{equation}
    \begin{gathered}
        \label{eq:XG2}
            Obj^{(t)}= \sum_{i=1}^{n} F_{loss}(y_{i},\hat{y_{i}}) + \sum_{k=1}^{t} F_{\lambda}(f_{k}), 
    \end{gathered}
\end{equation}
where $n$ is the number of predictions.

The loss function in XGBoost utilizes a second-order Taylor expansion.
As a result, the task of optimizing an objective function can be reduced to 
minimize a quadratic function.
In other words, one can evaluate the performance of the model by using the objective function following a specific node split in the decision tree.
If the performance of the decision tree model improves as a result of this node split, the modification will be implemented; otherwise, the node split will be terminated \cite{zhang2018data}. 
Because of the superior performance in supervised ML, the regressor model in this study is trained using the XGBoost algorithm. 
For the sake of brevity, further descriptions of the mathematical details of XGBoost are not provided here. For more details, see, e.g., Refs. \cite{purohit2022evaluation,chen2016xgboost}.
In this study, we have developed and utilized an in-house ML code using \textit{Python} version 3.8.3 to implement the ML algorithm utilized in the data-driven CFD simulations.

\subsection{Input features and output targets} \label{sec:IO}

The core objective, as stated in \autoref{sec:UQ md}, is to identify the corresponding location of the LES Reynolds stress anisotropy tensor on the barycentric map to establish the direction and magnitude of the perturbation.
As a result, when high-fidelity data is unavailable, we seek to predict the associated LES location.
The location of the LES point can be determined by predicting the horizontal distance ($\Delta \xi$) and vertical distance ($\Delta \eta$) between the corresponding points of RANS and LES, relying on baseline RANS values, as shown in \autoref{fig:Bary_tri}.
Consequently, $\Delta \xi$ and $\Delta \eta$ are variables that must be predicted using features from the baseline RANS simulation for each domain cell.
Since the target values are continuous variables, we are dealing with a regression model in which the mapping function between inputs and targets is determined using the XGBoost algorithm.

We use three categories of features as inputs for the regression model. The first category is chosen based on the explicit solution for the Reynolds stress anisotropy proposed by Pope \cite{Pope2000}. As given in \autoref{eq:pope}, the nonlinear turbulent-viscosity model is assumed to have a universal functional mapping between the strain-rate tensor $\mathbf{S} = 0.5 (\nabla \textbf{u} + (\nabla \textbf{u})^\text{T})$, the rotation-rate tensor $\mathbf{\Omega} = 0.5 (\nabla \textbf{u} - (\nabla \textbf{u})^\text{T})$, and the Reynolds stress anisotropy (i.e., $\mathbf{b}$), where $\nabla \textbf{u}$ represents the velocity-gradient tensor, and $\text{T}$ denotes the transpose of the matrix \cite{Pope2000},

\begin{equation}
    \begin{gathered}
        \label{eq:pope}
            \mathbf{b}(\mathbf{S},\mathbf{\Omega}) = \sum_{n=1}^{10} W^{(n)}(a_{i})L^{(n)}(\mathbf{S},\mathbf{\Omega}) = W^{(1)}(a_{i}) \mathbf{S} + W^{(2)}(a_{i}) (\mathbf{S}\mathbf{\Omega} - \mathbf{\Omega}\mathbf{S}) + W^{(3)}(a_{i})(\mathbf{S}^2 - \frac{1}{3}\text{trace}(\mathbf{S}^2)\mathbf{I}) + ...\ . 
    \end{gathered}
\end{equation}
Here, $W^{(n)}(a_{i})$ are scalar functions of the $a_{i}$ invariants, and $L^{(n)}$ are ten linearly independent basis tensors as a function of $\mathbf{S}$ and $\mathbf{\Omega}$. Further, by applying the Cayley-Hamilton theorem, it is possible to formulate various higher-order tensors as linear combinations of $L^{(n)}$ (see, e.g., Refs. \cite{Pope2000,wu2018physics, steiner2020data} for further details).   
There are two aspects of this assumption indicating that physics is missing. First, the pressure gradient affects turbulence such that it can be reduced by large values of favorable pressure gradient \cite{spalart2015philosophies}.
Second, the Reynolds stress at any point is solely determined by the local mean velocity $\textbf{u}(x,y,z)$ or its gradient $\nabla \textbf{u}(x,y,z)$, assuming that the production and dissipation of turbulence are balanced. However, the single-point-based turbulence constitutive equation is invalidated in many real-world applications because of strong non-equilibrium effects \cite{lumley1970toward}. 

We assume that the second category of features will make up for the lack of physical explanations outlined above. Hence, we include the turbulent kinetic energy gradient ($\nabla k$) and the pressure gradient ($\nabla p$)
in the input features,
resulting in a comprehensive functional mapping to the Reynolds stress from quantities of mean flow \cite{wu2018physics}. Since $\nabla p$ and $\nabla k$ are vectors, for convenience, they are transformed into their corresponding anti-symmetric tensors $\mathbf{G}_{p}$ and $\mathbf{G}_{k}$, respectively, as 

\begin{equation}
    \begin{gathered}
        \label{eq:antisymm}
            D_{jk} = 0.5\varepsilon_{ijk} d_{i}, 
    \end{gathered}
\end{equation}
where $D_{jk}$ is the anti-symmetric tensor form of vector $d_{i}$ and $\varepsilon_{ijk}$ represents the Levi-Civita tensor.
Using non-dimensional, rotationally invariant, and Galilean invariant features yields a generalizable model under rotation and Galilean transformations \cite{wu2018physics}.
Rotational invariance can be established using invariant features, as well as the norm of vector variables or the Frobenius norm of tensor variables. 
According to Galilean invariance, for constant velocity in all frames, the laws of motion would be the same.
We build a library of 47 features using four base tensors that must be normalized. \autoref{tab:norm} shows the normalized base tensors.

\begin{table}[ht]
\centering
    \caption{Base tensors normalization for input feature library. Normalization factors are  defined based on Refs. \cite{wu2018physics,yin2020feature}, where $\|.\|$ denotes the Frobenius norm of the matrix and  $|.|$ is the vector norm.}
    \label{tab:norm}
% \resizebox{\textwidth}{!}
{%
\begin{tabular}{@{}lll@{}}
\hline
Base tensor & Description                      & Normalized tensor \\ %\midrule
\hline
$\mathbf{S}$           & Rate of strain                   & $\hat {\mathbf{S}} = \mathbf{S}/(\|\mathbf{S}\|+ \varepsilon/k)$          \\
$\mathbf{\Omega}$       & Rate of rotation                 & $\hat {\mathbf{\Omega}}= \mathbf{\Omega}/(\|\mathbf{\Omega}\|+ \varepsilon/k)$     \\
$\mathbf{G}_{p}$          & Anti-symmetric pressure gradient & $\hat {\mathbf{G}}_{p}= \mathbf{G}_{p}/(\|{\mathbf{G}}_{p}\|+ \rho|(\bar{u}.\nabla)\bar{u}|)$      \\
$\mathbf{G}_{k}$           & Anti-symmetric turbulent kinetic energy gradient      & $\hat{\mathbf{G}}_{k}= \mathbf{G}_{k}/(\|{\mathbf{G}}_{k}\|+ \varepsilon/\sqrt{k})$       \\ %\bottomrule
\hline
\end{tabular}%
}
\end{table}

Wu et al. \cite{wu2018physics} assigned the input features as the first invariants of integrity basis tensors since they can not be utilized as input features directly. 
\autoref{tab:qs} lists 47 attributes %that have been 
proven to be helpful in ML and are utilized as a part of the input features in this study. 

\begin{table}[ht]
\centering
    \caption{Invariant basis list for input features (based on Ref. \cite{wu2018physics}). All terms created by cyclic permutation of anti-symmetric tensor labels are included when an asterisk appears next to a term, e.g., $\hat{\mathbf{G}}_{p}^2\hat{\mathbf{G}}_{k}\hat{\mathbf{S}}^*$ means $\hat{\mathbf{G}}_{p}^2\hat{\mathbf{G}}_{k}\hat{\mathbf{S}}$ and $\hat{\mathbf{G}}_{k}^2\hat{\mathbf{G}}_{p}\hat{\mathbf{S}}$. $\#s$ and $\#a$ denote symmetric and anti-symmetric normalized tensors' number.}
    \label{tab:qs}
{%
\begin{tabular}{lll}
\hline
(\#s, \#a) & Feature index    & Invariant bases                                      \\ \hline 
(1,0)  & $q_{1}$, $q_{2}$ & $\hat{\mathbf{S}}^2$, $\hat{\mathbf{S}}^3$  \\  
(0,1)  & $q_{3}$:$q_{5}$  & $\hat{\mathbf{\Omega}}^2$, $\hat{\mathbf{G}}_{p}^2$, $\hat{\mathbf{G}}_{k}^2$    \\ 
(1,1) &  $q_{6}$:$q_{14}$ & \begin{tabular}[c]{@{}c@{}}$\hat{\mathbf{\Omega}}^2\hat{\mathbf{S}}$, $\hat{\mathbf{\Omega}}^2\hat{\mathbf{S}}^2$, $\hat{\mathbf{\Omega}}^2\hat{\mathbf{S}}\hat{\mathbf{\Omega}}\hat{\mathbf{S}}^2$, 
$\hat{\mathbf{G}}_{p}^2\hat{\mathbf{S}}$, $\hat{\mathbf{G}}_{p}^2\hat{\mathbf{S}}^2$,  $\hat{\mathbf{G}}_{p}^2\hat{\mathbf{S}}\hat{\mathbf{G}}_{p}\hat{\mathbf{S}}^2$, 
$\hat{\mathbf{G}}_{k}^2\hat{\mathbf{S}}$, $\hat{\mathbf{G}}_{k}^2\hat{\mathbf{S}}^2$,  $\hat{\mathbf{G}}_{k}^2\hat{\mathbf{S}}\hat{\mathbf{G}}_{k}\hat{\mathbf{S}}^2$\end{tabular} \\ 
(0,2) & $q_{15}$:$q_{17}$ & $\hat{\mathbf{\Omega}}\hat{G}_{p}$, $\hat{\mathbf{G}}_{p}\hat{\mathbf{G}}_{k}$, $\hat{\mathbf{\Omega}}\hat{\mathbf{G}}_{k}$ \\ 
(1,2) &$q_{18}$:$q_{41}$ & \begin{tabular}[c]{@{}l@{}}$\hat{\mathbf{\Omega}}\hat{\mathbf{G}}_{p}\hat{\mathbf{S}}$, $\hat{\mathbf{\Omega}}\hat{\mathbf{G}}_{p}\hat{\mathbf{S}}^2$, $\hat{\mathbf{\Omega}}^2\hat{\mathbf{G}}_{p}\hat{\mathbf{S}}^*$, $\hat{\mathbf{\Omega}}^2\hat{\mathbf{G}}_{p}\hat{\mathbf{S}}^{2*}$, $\hat{\mathbf{\Omega}}^2\hat{\mathbf{S}}\hat{\mathbf{G}}_{p}\hat{\mathbf{S}}^{2*}$, 
$\hat{\mathbf{\Omega}}\hat{\mathbf{G}}_{k}\hat{\mathbf{S}}$, $\hat{\mathbf{\Omega}}\hat{\mathbf{G}}_{k}\hat{\mathbf{S}}^2$, $\hat{\mathbf{\Omega}}^2\hat{\mathbf{G}}_{k}\hat{\mathbf{S}}^*$,\\ $\hat{\mathbf{\Omega}}^2\hat{\mathbf{G}}_{k}\hat{\mathbf{S}}^{2*}$, $\hat{\mathbf{\Omega}}^2\hat{\mathbf{S}}\hat{\mathbf{G}}_{k}\hat{\mathbf{S}}^{2*}$,
$\hat{\mathbf{G}}_{p}\hat{\mathbf{G}}_{k}\hat{\mathbf{S}}$, $\hat{\mathbf{G}}_{p}\hat{\mathbf{G}}_{k}\hat{\mathbf{S}}^2$, $\hat{\mathbf{G}}_{p}^2\hat{\mathbf{G}}_{k}\hat{\mathbf{S}}^*$, $\hat{\mathbf{G}}_{p}^2\hat{\mathbf{G}}_{k}\hat{\mathbf{S}}^{2*}$, $\hat{\mathbf{G}}_{p}^2\hat{\mathbf{S}}\hat{\mathbf{G}}_{k}\hat{\mathbf{S}}^{2*}$\end{tabular} \\ 
(0,3) & $q_{42}$ & $\hat{\mathbf{\Omega}}\hat{\mathbf{G}}_{p}\hat{\mathbf{G}}_{k}$\\ 
(1,3) & $q_{43}$:$q_{47}$ & $\hat{\mathbf{\Omega}}\hat{\mathbf{G}}_{p}\hat{\mathbf{G}}_{k}\hat{\mathbf{S}}$,
$\hat{\mathbf{\Omega}}\hat{\mathbf{G}}_{k}\hat{G}_{p}\hat{\mathbf{S}}$,
$\hat{\mathbf{\Omega}}\hat{\mathbf{G}}_{p}\hat{\mathbf{G}}_{k}\hat{\mathbf{S}}^2$,
$\hat{\mathbf{\Omega}}\hat{\mathbf{G}}_{k}\hat{\mathbf{G}}_{p}\hat{\mathbf{S}}^2$,
$\hat{\mathbf{\Omega}}\hat{\mathbf{G}}_{p}\hat{\mathbf{S}}\hat{\mathbf{G}}_{k}\hat{\mathbf{S}}^2$\\ \hline
\end{tabular}%
}
\end{table}

Aside from the specified basis tensors and invariants, other physically interpretable features (the third category) listed in \autoref{tab:fs} can be added to account for the disparity between the RANS-predicted flow field and that from LES. All the aforementioned characteristics require physical intuition to connect physical trends to target values. Hence, they must have distinct physical meanings to ensure consistency in a flow with similar flow structures. $f_{1}$, $f_{2}$, and $f_{3}$ are based on Ling and Templeton's study \cite{ling2015evaluation}, $f_{4}$, $f_{6}$, and $f_{7}$ are inspired by the same study and used in the study of Tan et al. \cite{tan2021towards} as well. Finally, $f_{5}$ is included based on the study of Wang et al. \cite{wang2017comprehensive}. 
It should be noted that few features here may not be Galilean invariant due to the use of velocity in their formula (e.g., $f_{2}$). However, including those in the features' set is a common practice in similar studies, e.g., Ref. \cite{wu2018physics}.

\begin{table}[ht]
\centering
    \caption{The list of the physically-interpretable features.}
    \label{tab:fs}
{%
\begin{tabular}{@{}lll@{}}
%\toprule
\hline
Feature name & Description& Formulation \\ %\midrule
\hline
$f_{1}$ & Non-dimensionalized Q criterion & $(\|\mathbf{\Omega}\|^2-\|\mathbf{S}\|^2)/(\|\mathbf{\Omega}\|^2+\|\mathbf{S}\|^2)$          \\ 
$f_{2}$&Turbulence intensity& $k/(0.5\bar{u}_{i}\bar{u}_{i}+k)$          \\
$f_{3}$ & Ratio of turbulent timescale to mean strain time scale & $\|\mathbf{S}\|k/(\|\mathbf{S}\|k+\varepsilon)$          \\
$f_{4}$ & Viscosity ratio & $\nu_{t}/(500\nu+\nu_{t})$          \\
$f_{5}$   &Ratio of the total to normal Reynolds stress& $\|\bar{R}_{ij}\|/(2k)$          \\
$f_{6}$   &Normalized strain vs. vorticity I& $\tanh({\|\mathbf{S}\|/\|\mathbf{\Omega}\|})$          \\
$f_{7}$& Normalized strain vs. vorticity II& $\tanh({(\|\mathbf{\Omega}\|^2-\|\mathbf{S}\|\|\mathbf{\Omega}\|)/\|\mathbf{S}\|^2})$     
\\ %\bottomrule
\hline
\end{tabular}%
}
\end{table}
There also exist several interpretable physical properties that are used in other studies (e.g., in Refs. \cite{heyse2021estimating, zhu2019machine}). Most of them are equivalent to the parameters given in \autoref{tab:fs} with different mathematical expressions, and the remaining that are not consistent with the physics of the problem under investigation are not considered here.

\section{Results and Discussion} \label{sec:R&D}
In this section, first, we provide the results of the data-driven and data-free approaches for Case A. Then, a two-step feature-selection algorithm is described, and the final features are introduced. Tuning of the hyper-parameters and descriptions of the ML model cross-validation procedure are presented afterward. Finally, predictions made by the ML model for the unseen Cases B and C are thoroughly discussed. 
\subsection{Data-free vs. data-driven approach} \label{sec:predict}
\begin{figure}[ht]
    \centering
    \includegraphics[width=0.7\linewidth]{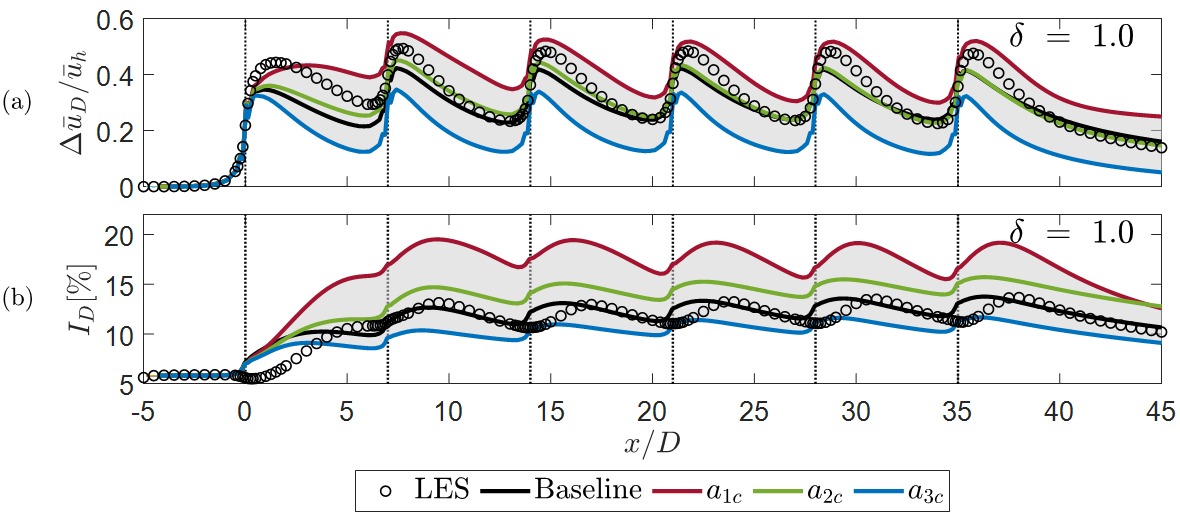}
    \caption{Rotor-averaged normalized velocity deficit (a) and turbulence intensity (b) in Case A based on the data-free approach for $\delta$ = 1.0. }
    \label{fig:df_rotor}
\end{figure}

In the data-free approach, there is no $a$ $priori$ knowledge about the perturbation value and direction, and commonly a constant $\delta$ is applied to the entire domain. In the most conservative case, $\delta$ = 1.0 is used for perturbation covering the LES data. This is shown in \autoref{fig:df_rotor} for the normalized velocity deficit and turbulence intensity; both averaged across the rotor area in a hypothetical cylinder starting from the inlet in Case A.
The baseline RANS results show a faster wake recovery for all turbine wakes indicated by the less normalized velocity deficit, which occurs more intensely for the most upstream turbine.
Perturbation toward $a_{1c}$ amplifies the shear and weakens the entrainment of momentum into the wake region, and consequently results in a reduced wake recovery compared to the LES results.
When the anisotropy tensor is perturbed toward the isotropic state (i.e., $a_{3c}$), due to the enhanced level of mixing, the wake recovery is overestimated \cite{eidi2021model}.
The perturbed values toward $a_{2c}$ lie within the bound and do not affect the uncertain quantity as discussed in \autoref{sec:UQ md}. As shown in this figure, utilizing a constant perturbation value toward all the limiting states of turbulence leads to a wide bound when attempting to cover the high-fidelity results (i.e., the gray area in \autoref{fig:df_rotor}) and, consequently, high uncertainty in estimating the QoIs. 
The reader may refer to our prior study \cite{eidi2021model} for further details on the data-free approach.

Using a data-driven approach, one can utilize a $priori$ determined perturbation value in the direction in which perturbation is necessary. For instance, when $d_{1c, LES} < d_{1c, RANS}$ (see \autoref{fig:Bary_tri}), perturbing RANS toward $a_{3c}$ expands the bandwidth and estimates the perturbed model's outputs farther from the high-fidelity value (i.e., increasing uncertainty and decreasing accuracy). On the other hand, for the same point, the one-component perturbation shifts QoIs toward the high-fidelity output and covers it. 
In \autoref{tab:delta}, the statistical results for the perturbation direction and amount are given for Case A, based on our data-driven framework. It is worth highlighting that only 1\% of the points have to be perturbed toward the two-component corner, which is another reason to assert that perturbing points toward $a_{2c}$ would not affect the uncertainty level. Moreover, the standard deviation values of $\delta$ represent the scatter distribution of perturbation values in the domain. 
This also verifies the usage of point-specific $\delta$ quantities rather than constant values. 

\begin{table}[ht]
\centering
    \caption{Perturbation direction based on the study framework and statistical details of $\delta$ for wake-zone points in Case A. Here, 10m-spacing upstream and downstream of the turbines are excluded.}
    \label{tab:delta}
{%
\begin{tabular}{@{}llllll@{}}
%\toprule
\hline
Perturbation direction & Percentage of the points to be perturbed & Mean $\delta$ & Standard deviation of $\delta$\\ %\midrule
\hline
$a_{1c}$               & 91\%    & 0.35   & 0.13        \\
$a_{2c}$               & 1\%     & 0.32   & 0.17        \\
$a_{3c}$               & 8\%     & 0.34   & 0.21         \\% \bottomrule
\hline
\end{tabular}%
}
\end{table}

\begin{figure}[ht]
    \centering
    \includegraphics[width=0.35\linewidth]{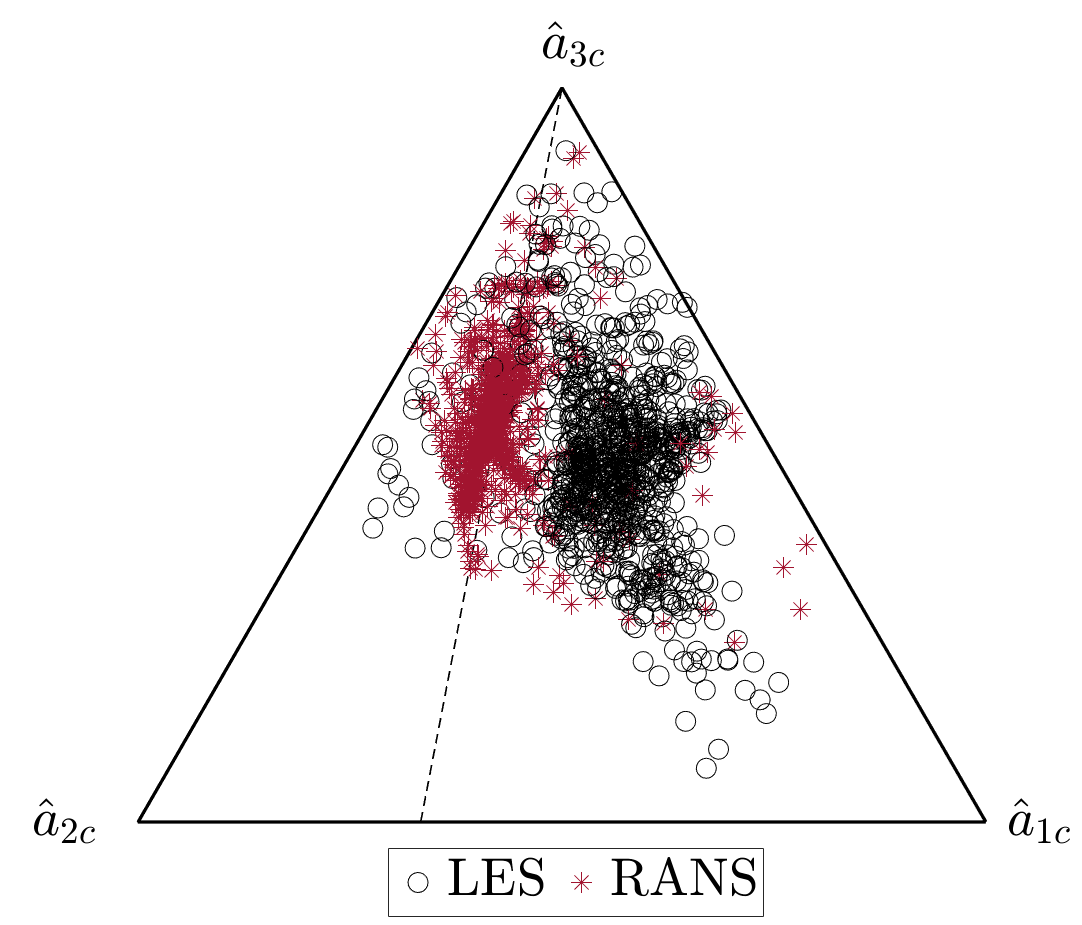}
    \caption{The LES and RANS Reynolds stress anisotropy shown on the barycentric map for approximately 1000 randomly picked points from the wake zone in Case A. Here, 10 m-spacing upstream and downstream of the turbines are excluded.}
    \label{fig:Bary_anisotropy}
\end{figure}

In \autoref{fig:Bary_anisotropy}, the RANS and LES Reynolds stress anisotropy for Case A are depicted on the barycentric map, where approximately 1000 points are randomly picked from the wake zone.
The results in this figure are entirely consistent with the statistics presented in \autoref{tab:delta}. The RANS points tend to be more isotropic than the LES data. Hence, most of the points need to be perturbed toward $a_{1c}$. There are very few points in which RANS is closer to the two-component corner on the map than LES; therefore, perturbing these few points toward $a_{2c}$ would have a negligible effect on the results.

\begin{figure}[ht]
    \centering
    \includegraphics[width=0.7\linewidth]{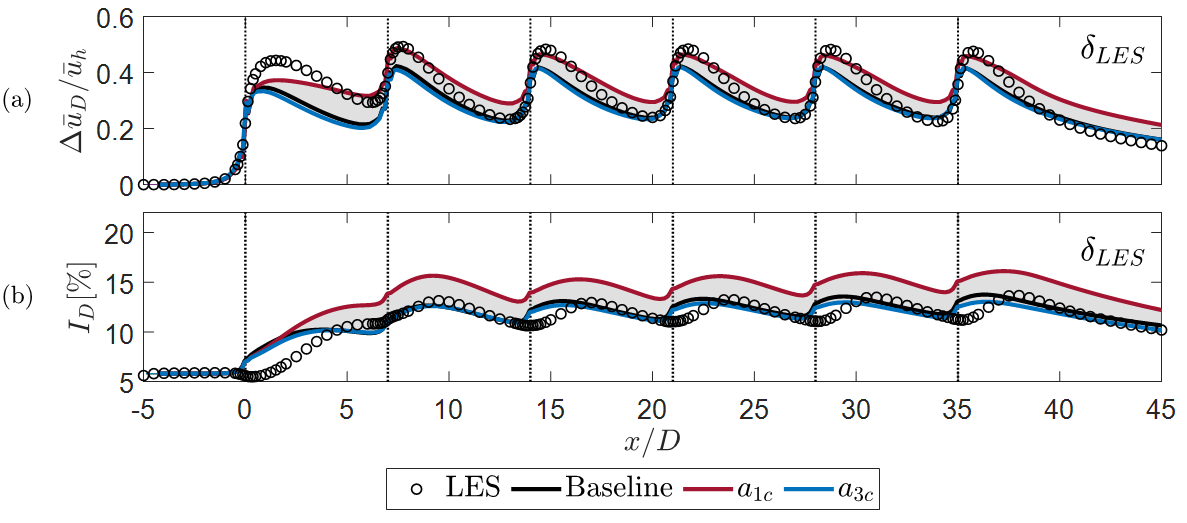}
    \caption{Rotor-averaged normalized velocity deficit (a) and turbulence intensity (b), with $\delta$ calculated based on the LES data.}
    \label{fig:dd_rotor}
\end{figure}

For the individual cells in the wake zone, utilizing the $\delta$ from the data-driven method with perturbation in a single direction ($a_{1c}$ or $a_{3c}$) tends to a narrower bound, capturing the LES results with higher accuracy. 
For case A, this is shown in \autoref{fig:dd_rotor} for the normalized velocity deficit and turbulence intensity; both averaged across the rotor area.

\begin{figure}[ht]
    \centering
    \includegraphics[width=1.0\linewidth]{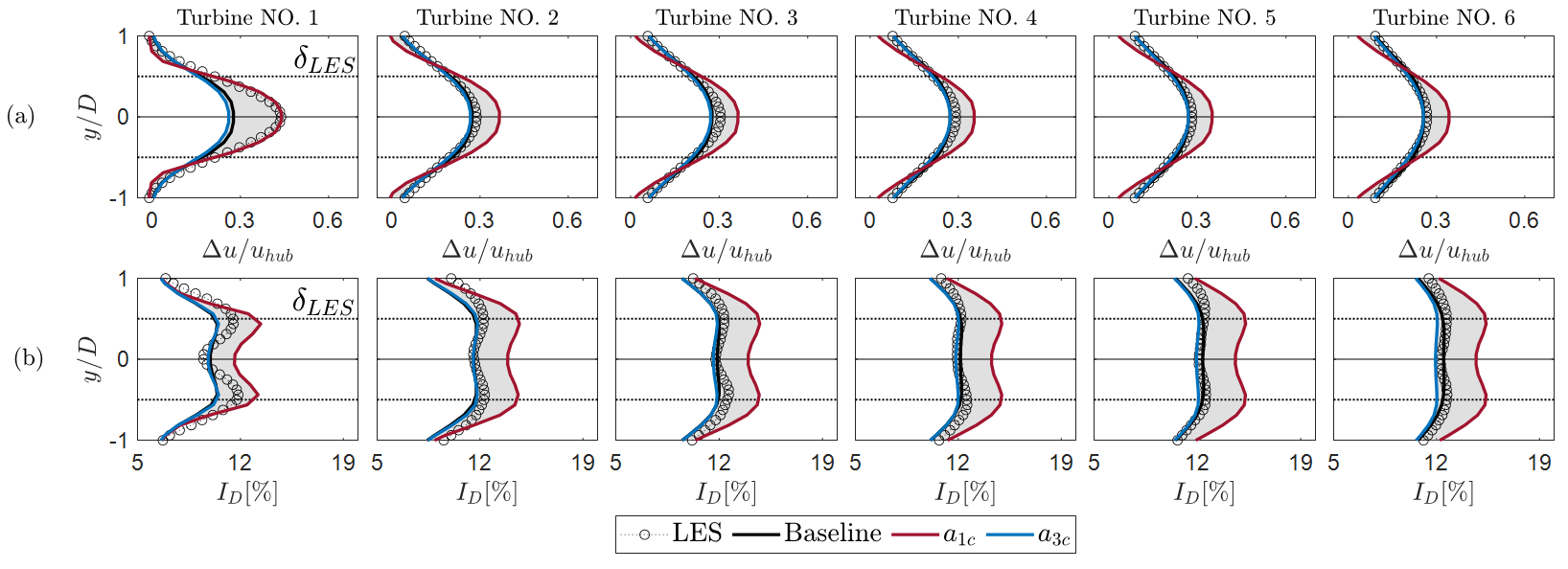}
    \caption{Lateral profiles of the normalized velocity deficit (a) and turbulence intensity (b) at $5D$ downstream of each turbine at the hub height in Case A. Here, $\delta$ calculated based on the LES data are utilized for perturbing RANS toward one- and three-component turbulence.}
    \label{fig:dd_NVD_TI_5D}
\end{figure}

\autoref{fig:dd_NVD_TI_5D} depicts the lateral profiles of the same quantities on a horizontal plane at $5D$ downstream distance of each turbine at $z_{h}$. Based on the isotropic eddy-viscosity hypothesis in the RANS models and the data-driven framework of the current study, we may claim that perturbation toward $a_{1c}$ uses larger $\delta$ values leading to a wider bound on the one-component perturbation side. On the other hand, perturbed results toward $a_{3c}$ do not significantly alter QoIs, indicating smaller $\delta$ values and fewer points that need to be perturbed.

Comparing the results of the data-free and data-driven frameworks reveals that the determination of perturbation value based on the data obtained from LES can significantly improve the UQ and leads to a more optimal bound over the LES data.
There exist an error in estimating uncertainty bound using both methods for the most upstream turbine's wake. This could be related to the deficiency of the eigenvalue perturbation method in this area.
However, since the LES values, excluding the turbines' vicinity, are well covered in the wake region with more accuracy and less uncertainty, the proposed data-driven approach outperforms the constant $\delta$ method.

\subsection{Feature selection} \label{sec:FS}
As discussed in \autoref{sec:IO}, 54 features are chosen as the inputs of the regression model, which is very likely to cause over-fitting and high computational cost.
A feature can be relevant, which affects the target variable and should be retained, or it is irrelevant, and its effect on the target variable is zero or negligible. A feature can also be redundant, meaning that its impact on the target variable is equivalent to the effect of another feature. In the feature-selection phase, irrelevant and redundant features should be discarded.

Defining the interdependency of variables is a typical method for selecting the essential features that are most dependent on the output variable and have the least dependency on the other inputs.
As one of the feature-selection methods, the filter approaches are based on statistical tests or metrics from the information theory. They have the advantage of being computationally cost-effective, allowing many features to be examined fast. For instance, the correlation coefficient can find linear and monotonic relationships between the target variables and the features. Mutual information (MI) can be used to evaluate a broader reliance and is a metric that assesses how much information is shared between two variables \cite{thomas2006elements}.

The MI between continuous arbitrary variables $X$ and $Y$ can be estimated by 

\begin{equation}
    \begin{gathered}
        \label{eq:MI}
            MI(X;Y)=\int_{Y} \int_{X}p_{(X,Y)}(x,y)\log \frac{p_{(X,Y)}(x,y)}{p_{X}(x)p_{Y}(y)} \,dx\,dy\,, 
    \end{gathered}
\end{equation}
where $p_{(X,Y)}(x,y)$ denotes the joint probability density between the variables, and $p_{X}(x)$ and $p_{Y}(y)$ represent marginal distributions. The difference between the joint distribution and its marginal distributions is calculated by utilizing the MI. By using the entropy concept and denoting the average level of uncertainty in the outcome of the random variable \cite{steiner2020data}, the MI can also be expressed as

\begin{equation}
    \begin{gathered}
        \label{eq:entropy}
            MI(X;Y) = e(X)- e(X|Y) = e(X) + e(Y) - e(X,Y),     \end{gathered}
\end{equation}
where $e(X) = -\int_{X}p_{X}(x)\log p_{X}(x) \,dx\,$, $e(X|Y)$,
and $e(X,Y)$ are differential, conditional differential, and point differential entropy, respectively. As a result, the MI can be defined as the amount of uncertainty in $X$ that is eliminated due to knowing $Y$. In other words, when $Y$ is unrelated to $X$, $e(X|Y) = e(X)$, and consequently $MI(X;Y) = 0$.
In this study, we estimate entropy using the $k$-nearest neighbor distance as a cost-effective method (see, e.g., Ref. \cite{ ross2014mutual} for further details).

\begin{figure}[ht]
    \centering
    \includegraphics[width=0.9\linewidth]{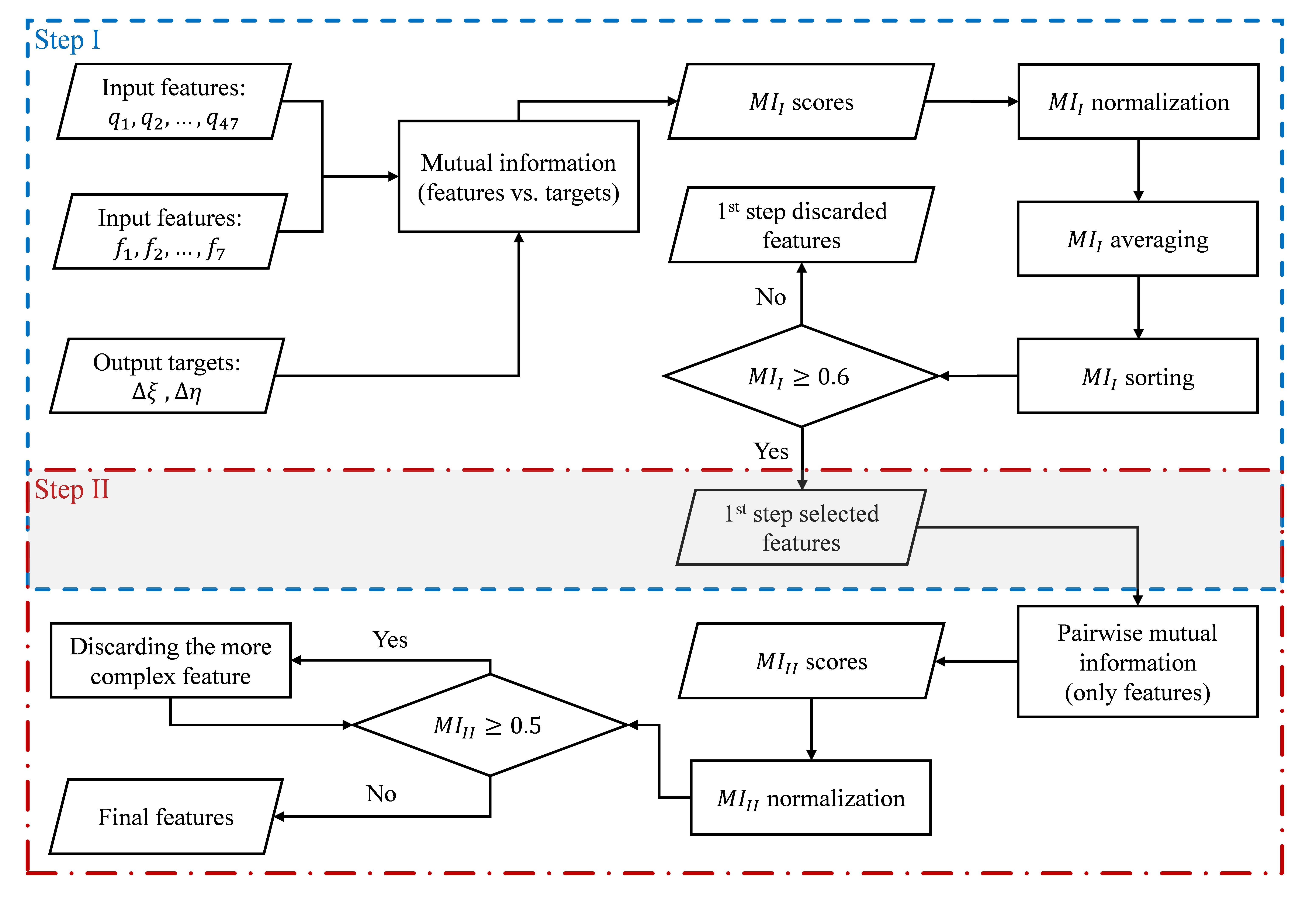}
    \caption{Flowchart of the two-step feature selection based on the mutual information metric. \black}
    \label{fig:feature_selection}
\end{figure}

\begin{figure}[ht]
     \centering
     \includegraphics[width=0.6\linewidth]{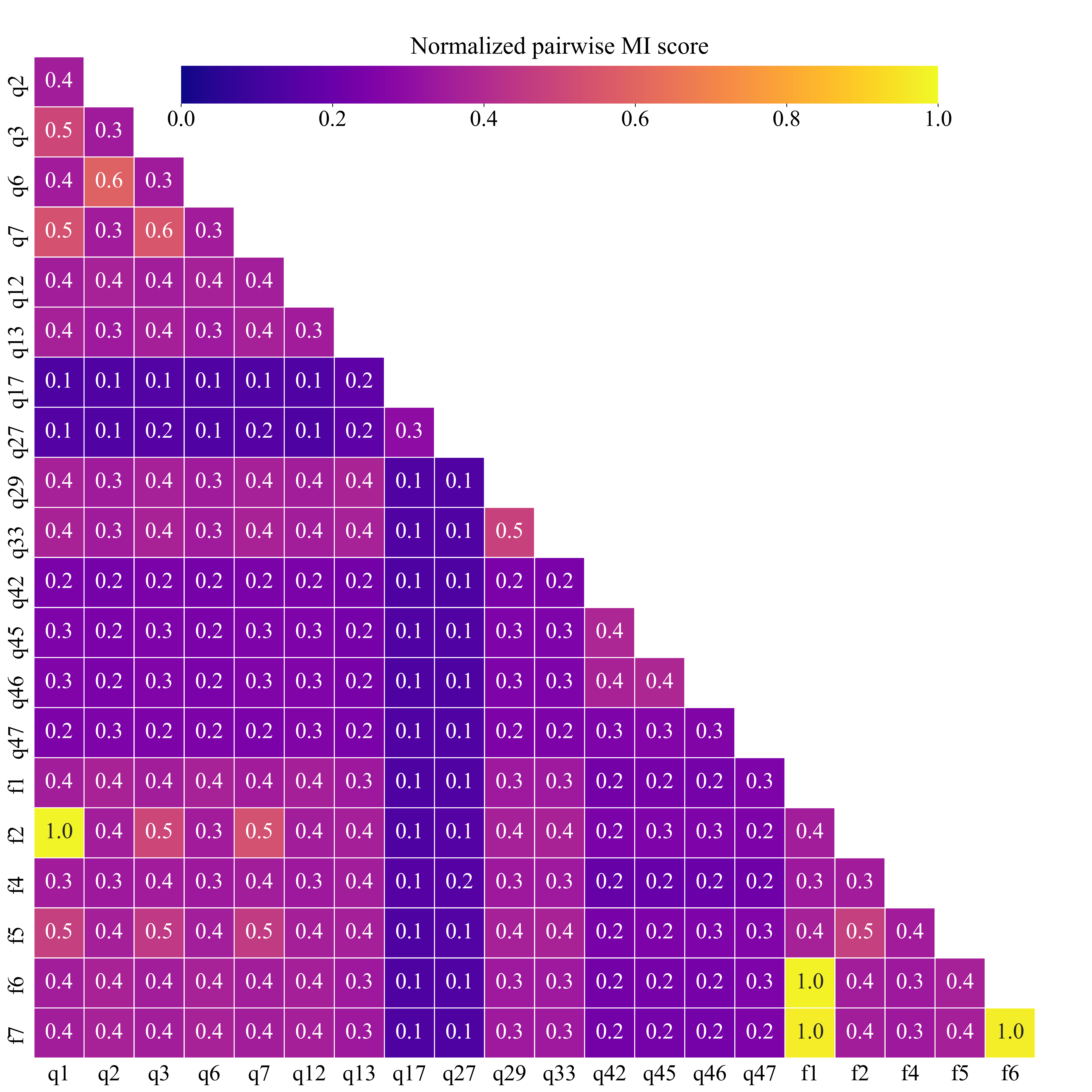}
     \caption{The pairwise MI scores for 21 features selected after the first step of the feature-selection procedure.}
     \label{fig:MI_score}
 \end{figure}

In the current study, we apply the MI (see, \code{mutual info regression} in Ref. \cite{scikit-learn}) in two steps as shown in \autoref{fig:feature_selection}.
First, the MI between 54 features and their corresponding two targets for each domain cell is calculated, normalized, averaged, and sorted ($MI_{I}$). Then, with the criterion of $MI_{I} \geq$ 0.6, 33 irrelevant features are eliminated. 
In the second step, a pairwise MI ($MI_{II}$) is determined among the remaining 21 features to distinguish the redundant ones. \autoref{fig:MI_score} indicates the pairwise normalized rounded MI score values for the remaining features. The more complex features are removed when the normalized pairwise $MI_{II} \geq$ 0.5. Finally, 13 features out of the initial 54 features are chosen as the inputs of the regression model as  

\begin{equation}
    \begin{gathered}
        \label{eq:set}
           \text{Final input features set} = \{q_{1}, q_{2}, q_{12}, q_{13}, q_{17}, q_{27}, q_{29}, q_{42}, q_{45}, q_{46}, q_{47}, f_{1}, f_{4}\}.  
    \end{gathered}
\end{equation}   

\subsection{Validation of the machine learning model} \label{sec:MT}
In this study, we use Case A as the baseline for training and testing the ML model, and the verified ML model is utilized to predict QoIs for the other two cases.
In other words, by having the LES Reynolds stress anisotropy for Case A, the exact location of the corresponding LES point on the barycentric map is known.
Then, for Case A, $\Delta \xi$ and $\Delta \eta$ values can be used as the model labels; however, for Cases B and C, we presume that the target values are unknown and will be predicted by the ML model.
The predicted values are then compared to the LES results for the two unseen cases to assess the accuracy of the ML model. 

There exist several hyper-parameters related to the architecture of the XGBoost model that the user must tune along with the automatically computed parameters that are updated during the training process (for a detailed description of the hyper-parameters, see, e.g., Ref. \cite{scikit-learn}).
In this study, only five hyper-parameters of XGBoost are considered.
We use the randomized search \cite{bergstra2012random} and the grid search algorithms to determine optimal values for the hyper-parameters.
First, the appropriate intervals of the hyper-parameters are assumed considering their distribution, and the approximate optimal values are determined by the randomized search algorithm.
Then, the grid search algorithm calculates the relatively tuned hyper-parameters considering a set of rounded values based on the randomized search results.
The tuned hyper-parameters are given in \autoref{tab:hyper}.
We use repeated ten-fold cross-validation for the model's training process, and the coefficient of determination ($R^2$, see, $\code{r2 score}$ in \cite{scikit-learn}) is utilized as the model's performance measure.

\begin{table}[ht]
\centering
    \caption{Tuned hyper-parameter of the XGBoost regression model.}
    \label{tab:hyper}
{%
\begin{tabular}{@{}ll@{}}
%\toprule
\hline
Hyper-parameter    & Optimized value \\% \midrule
\hline
Booster            & gbtree          \\
Number of trees    & 500             \\
Maximum tree depth & 5               \\
% Gamma              & 0               \\
Learning rate      & 0.2             \\
Sub-sample ratio   & 0.8             \\ %\bottomrule
\hline
\end{tabular}%
}
\end{table}

It is important to note that two other well-known ML algorithms, including random forest \cite{breiman2001random} and support vector machine \cite{drucker1996linear}, are investigated along with the XGBoost algorithm.
The XGBoost algorithm offers a higher accuracy ($R^2$ = 0.98 in cross-validation) with a significantly reduced computational time compared to the other two. The $R^2$ scores of the random forest and support vector machine algorithms are equal to 0.93 and 0.94 in the training phase, respectively. As a result, the XGBoost method is selected as the main regression model for this study.

\subsection{Predictions of machine learning model} \label{sec:Mp}
\begin{figure}[htb]
    \centering
    \includegraphics[width=0.7\linewidth]{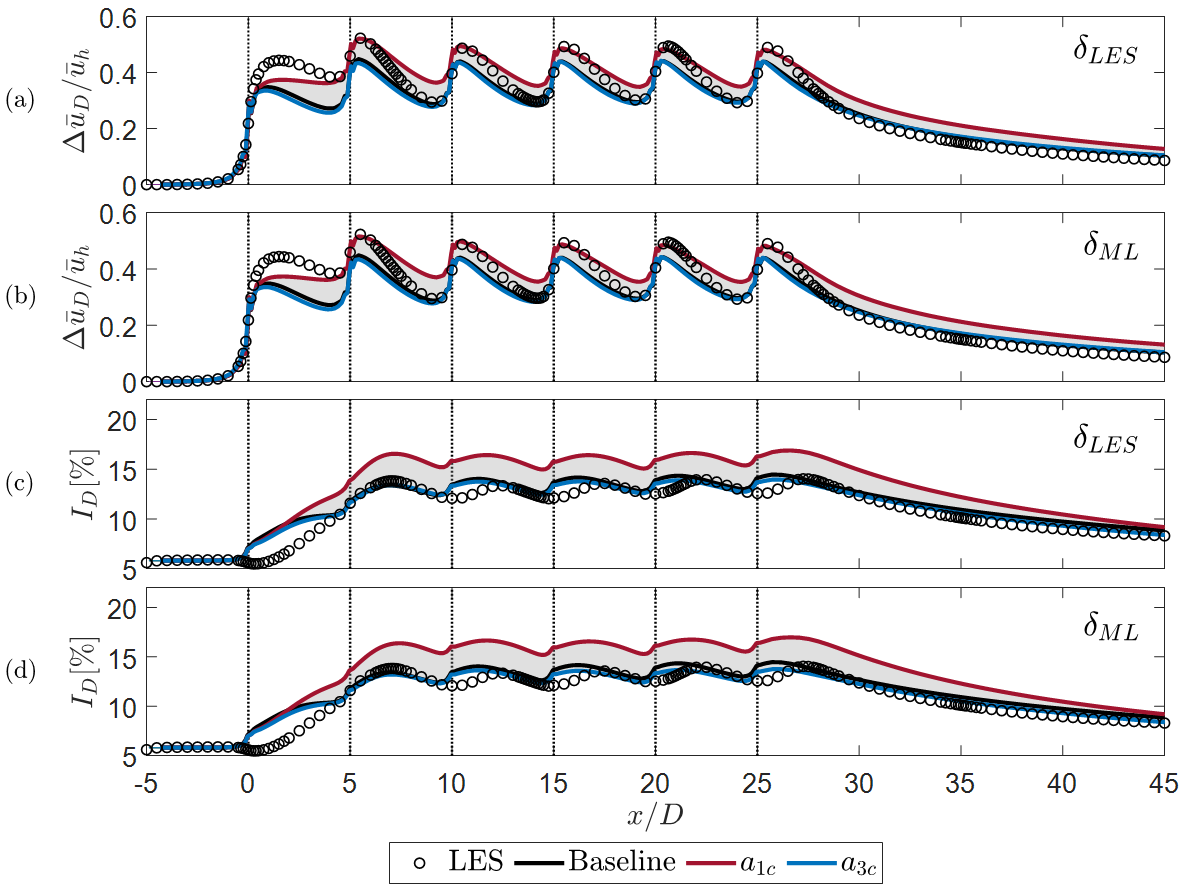}
    \caption{Rotor-averaged normalized velocity deficit (a,b) and turbulence intensity (c,d) in Case B. Here, for the perturbed RANS (toward one- and three-component turbulence), $\delta$ calculated based on LES data (a,c) and $\delta$ predicted by the ML model (b,d) are utilized.}
    \label{fig:dd_NVD_TI_rotor_B}
\end{figure}

To assess the ML model's performance for the unseen cases, the trained model predicts $\Delta \xi$ and $\Delta \eta$ values for the unlabeled input data of Case B and Case C.
As a result, for each point in the wake area in these two cases, the estimated location of the LES Reynolds stress anisotropy on the barycentric map is determined.
\autoref{fig:dd_NVD_TI_rotor_B} indicates the results of the perturbed model using $\delta$ calculated from LES values and ML-predicted $\delta$, both for Case B. 
Similar to Case A, the high-fidelity results are well captured by the uncertainty bound in the distant wake.
However, similar to the data-free results, a discrepancy exists in the near-wake area due to the deficiencies of the baseline model in the vicinity of the turbines.
This error in the UQ based on the suggested framework appears to be insignificant since the downstream turbines in wind farms are not usually positioned in the very near-wake region. 

\begin{figure}[ht]
    \centering
    \includegraphics[width=1.0\linewidth]{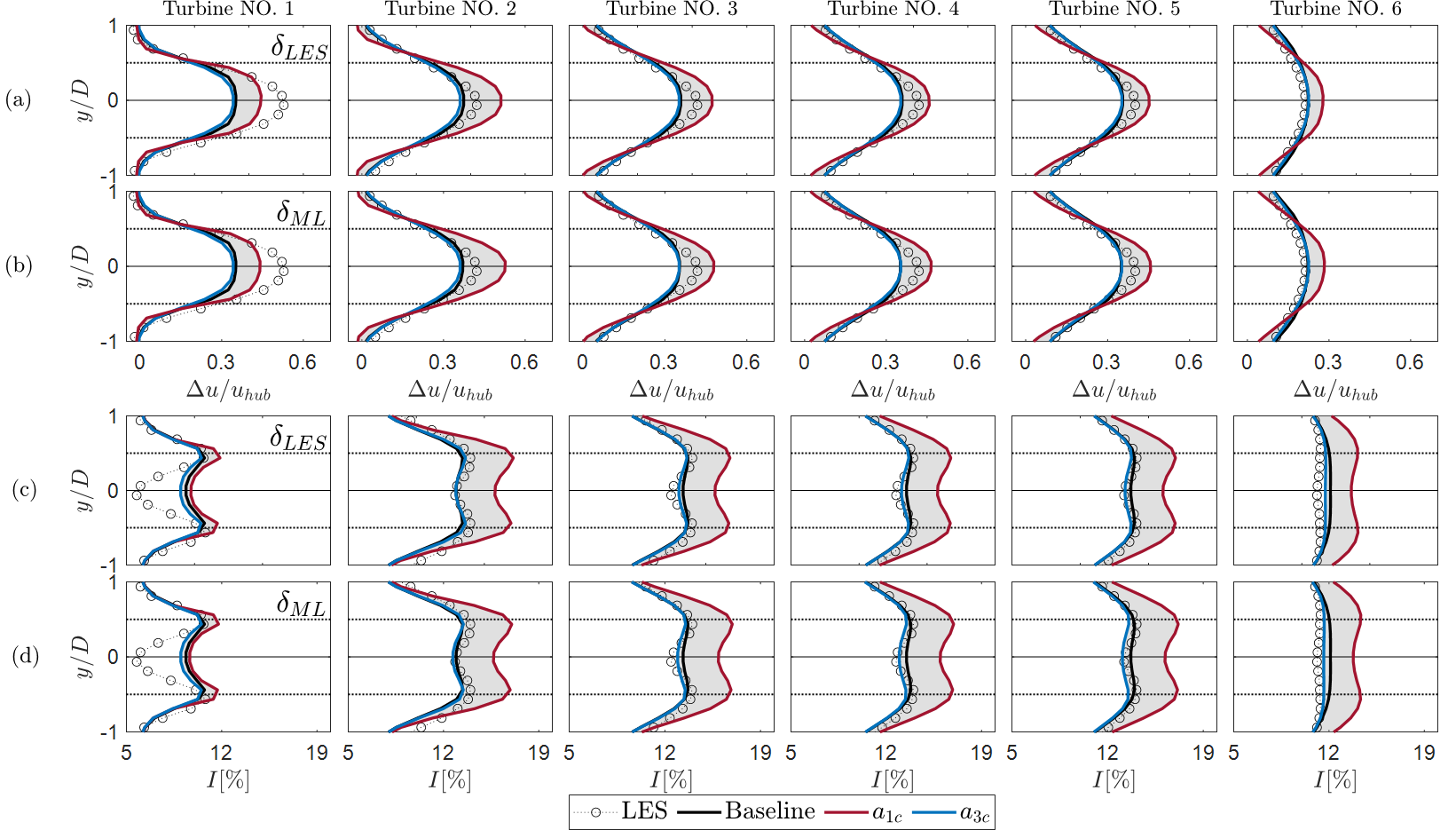}
    \caption{Lateral profiles of the normalized velocity deficit (a,b) and turbulence intensity (c,d) at $3D$ downstream of each turbine at the hub height in Case B. Here, for the perturbed RANS (toward one- and three-component turbulence), $\delta$ calculated based on LES data (a,c) and $\delta$ predicted by the ML model (b,d) are utilized.}
    \label{fig:dd_NVD_TI_3D_B}
\end{figure}

\autoref{fig:dd_NVD_TI_3D_B} depicts the lateral profile of the normalized velocity deficit and turbulence intensity at a $3D$ distance downstream of turbines at the hub height in Case B. The results show that, except for the most upstream turbine, the ML model can accurately predict the LES Reynolds anisotropy location for individual cells even when the turbines' spacing in the wind farm is changed. Considering the fact that in Case B, the ML algorithm determines $\delta$ values for the wake-area perturbation solely based on the input features, the ML-predicted results for this unseen case seem almost identical to those obtained using the LES anisotropy to calculate the $\delta$ values. 

\begin{figure}[ht]
    \centering
    \includegraphics[width=1.0\linewidth]{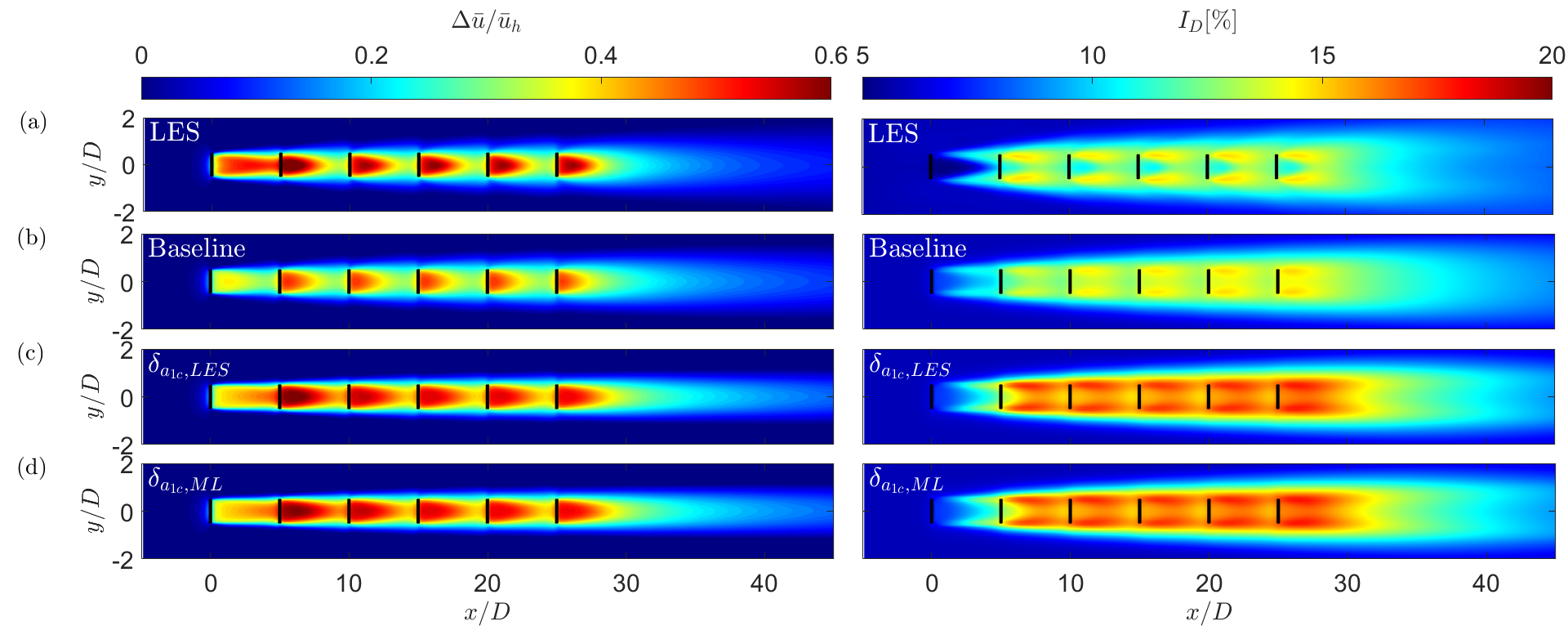}
    \caption{Contours of normalized velocity deficit (left) and turbulence intensity (right) for LES (a), baseline RANS (b), and data-driven perturbed models toward $a_{1c}$ with $\delta$ calculated based on LES data (c) and $\delta$ predicted by the ML model (d) in Case B.}
    \label{fig:b_contour}
\end{figure}

To further explore the flow physics and evaluate the results in comparison with the baseline model and high-fidelity data, contours of normalized velocity deficit and turbulence intensity are displayed at the hub height for Case B in \autoref{fig:b_contour}.
As discussed in the prior figures, the results of RANS, perturbed towards $a_{3c}$, are close to the baseline; therefore, contour plots for the isotropic perturbation are not given here.
The normalized velocity deficit contours show a faster wake recovery in the baseline model than that in LES, resulting in prediction of higher velocity values compared to the LES data.
Perturbation toward $a_{1c}$ provides a more accurate estimation of the velocity in the wake region and increases the velocity deficit.
Turning to the turbulence intensity contour plots, the baseline model has a relatively good estimation compared to LES, and perturbation toward $a_{1c}$ will increase the shear and, consequently, the turbulence intensity behind the turbines.
The figure also reveals that the trained ML model is accurate when applied for the unseen cases since the prediction results for both QoIs are in good agreement with those obtained by the data-driven framework based on LES data. 

\begin{figure}[ht]
    \centering
    \includegraphics[width=0.7\linewidth]{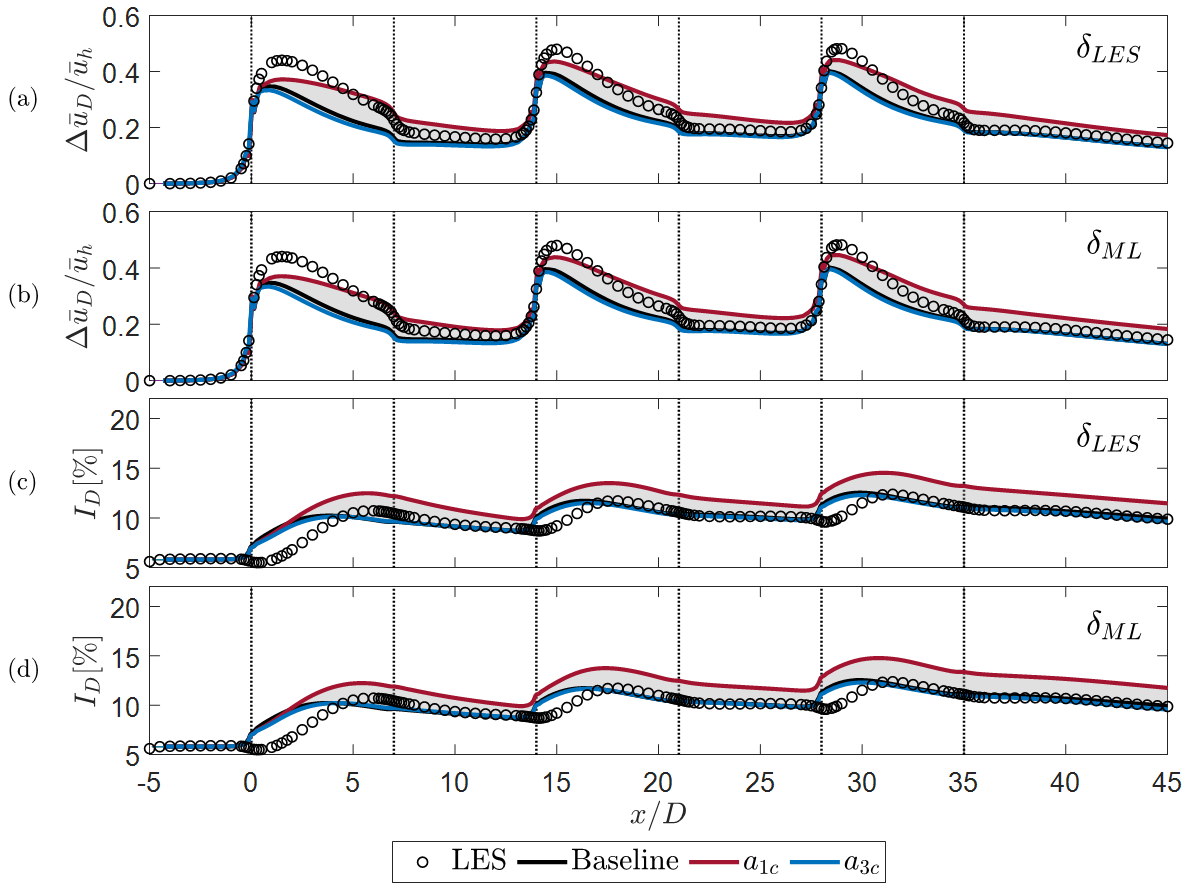}

    \caption{Normalized velocity deficit (a,b) and turbulence intensity (c,d) averaged across the rotor area of odd turbine rows in Case C. Here, for the perturbed RANS (toward one- and three-component turbulence), $\delta$ calculated based on LES data (a,c) and $\delta$ predicted by the ML model (b,d) are utilized.}
    \label{fig:dd_NVD_TI_rotor_C}
\end{figure}

\begin{figure}[ht]
    \centering
    \includegraphics[width=1.0\linewidth]{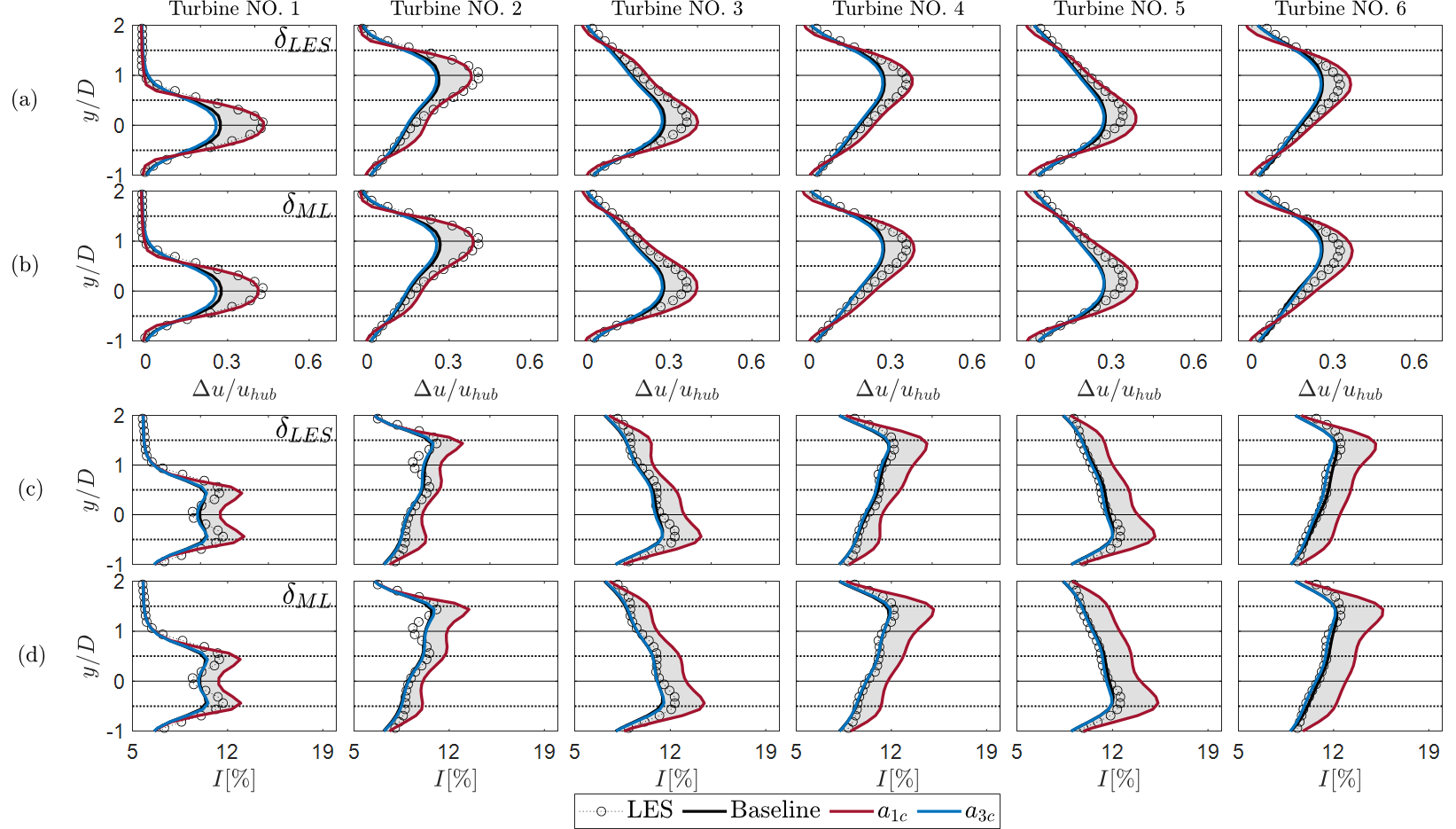}
    \caption{Lateral profiles of the normalized velocity deficit (a,b) and turbulence intensity (c,d) at $5D$ downstream of each turbine at the hub height in Case C.  Here, for the perturbed RANS (toward one- and three-component turbulence), $\delta$ calculated based on LES data (a,c) and $\delta$ predicted by the ML model (b,d) are utilized.}
    \label{fig:dd_NVD_TI_5D_C}
\end{figure}

To examine the ability of the ML model to predict the target values for a farm with a not fully-aligned layout, the QoIs are shown in \autoref{fig:dd_NVD_TI_rotor_C} for Case C, averaged across the rotor area of the odd turbine rows.
The lateral profiles of the same QoIs are shown in \autoref{fig:dd_NVD_TI_5D_C} at a streamwise location of $5D$ downstream of the turbines at the hub height.
As can be observed, all turbines' wakes have accurate bandwidth prediction and optimal coverage of the high-fidelity results.
The accuracy of the results in Case C demonstrates the validity of the UQ framework in a more complicated wind-farm configuration, as well as the ML model's independency from the turbine alignment.

\begin{figure}[ht]
    \centering
    \includegraphics[width=1.0\linewidth]{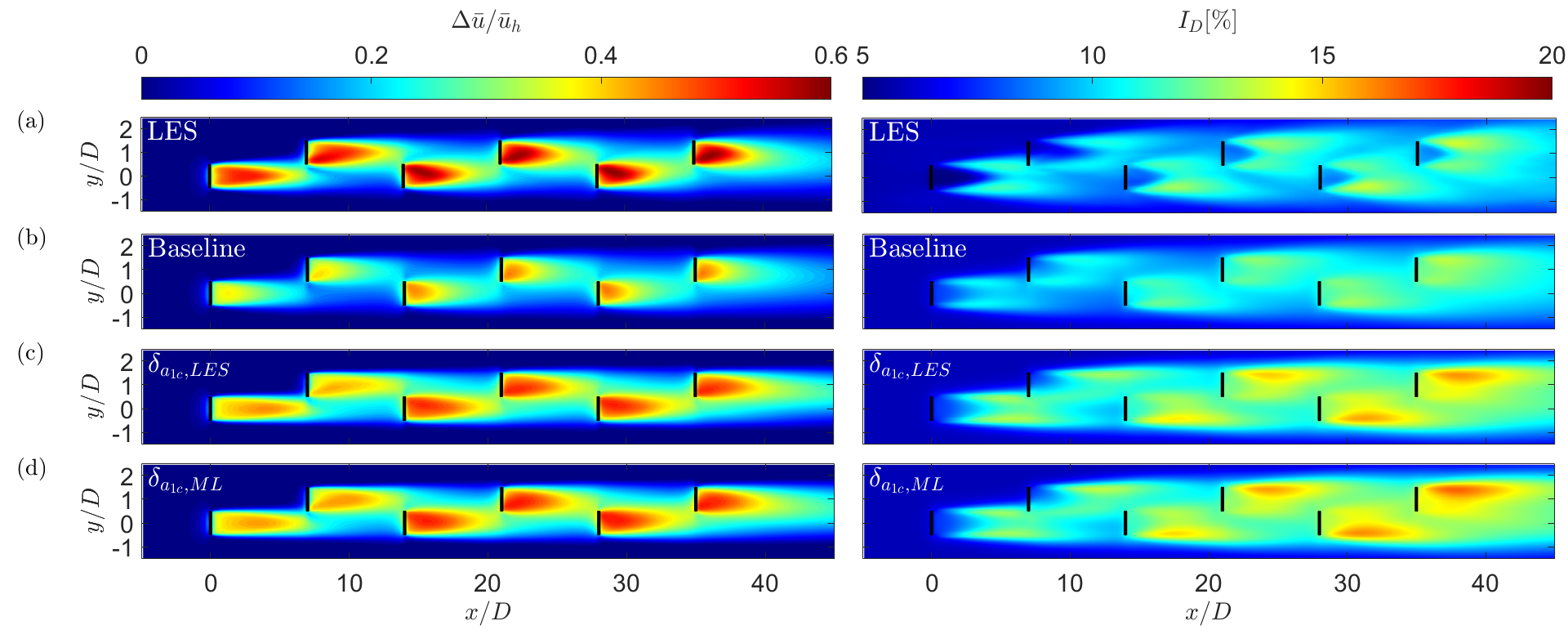}
    \caption{Contours of normalized velocity deficit (left) and turbulence intensity (right) for LES (a), baseline RANS (b), and data-driven perturbed models toward $a_{1c}$ with $\delta$ calculated based on LES data (c) and $\delta$ predicted by the ML model (d) in Case C.}
    \label{fig:c_contour}
\end{figure}

The contour plots of the normalized velocity deficit and turbulence intensity at the hub height for Case C are depicted in \autoref{fig:c_contour}.
Compared to the other two layouts, the non-alignment of the turbines results in the development of partial wake and a more complex flow.
As a result, the baseline model indicates a weaker velocity estimation.
Hence, the wake asymmetry appears to be more severe in the LES compared to the baseline model.
The perturbed model toward $a_{1c}$ largely resolves this problem.
Results of perturbation toward $a_{3c}$ are excluded in this case either.
Similar to Case B, the baseline model gives relatively accurate estimates of the turbulence intensity at the hub height, but perturbation toward anisotropic turbulence overestimates the outputs.
However, the overall results of both perturbed models would make a proper coverage over LES data.
ML-predicted QoIs are almost identical to the data-driven LES-informed results through the framework of this study.

\begin{figure}[ht]
    \centering
    \includegraphics[width=1.0\linewidth]{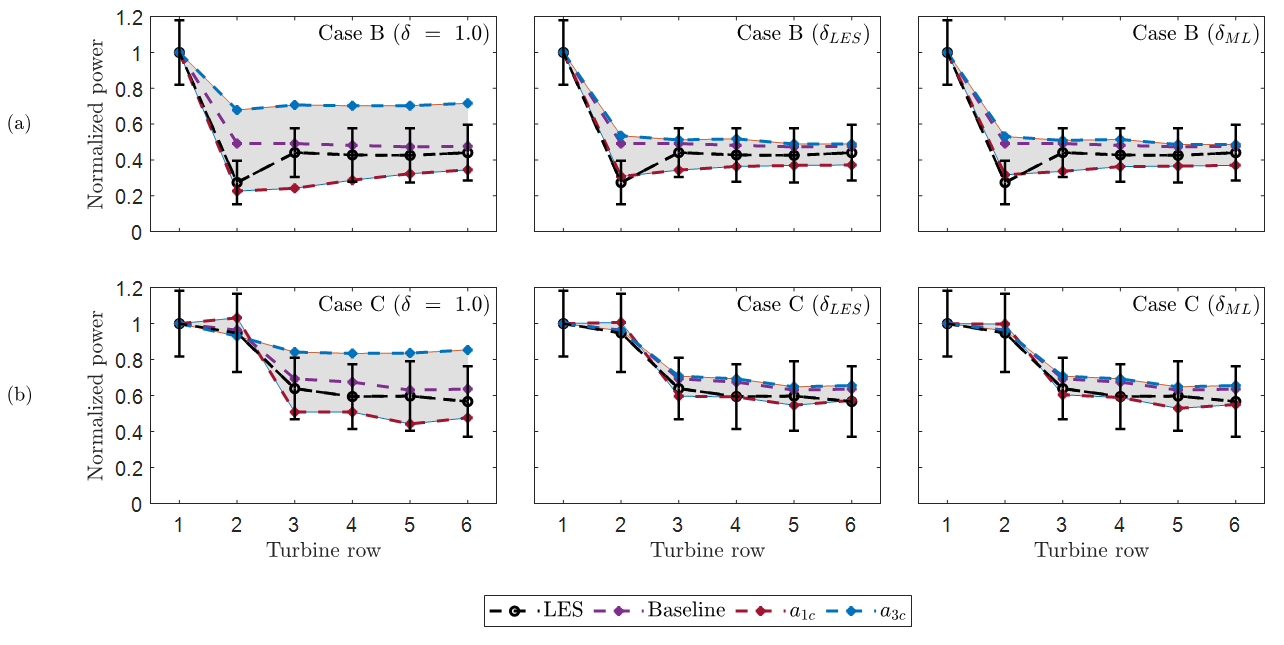}
    \caption{Normalized power output for turbine rows in Case B (a) and Case C (b), based on the data-free approach with $\delta$= 1.0 (left column), the data-driven approach with $\delta$ calculated from LES data (middle column), and the data-driven approach with $\delta$ predicted by the ML model (right column). The vertical bar indicates the standard deviation of the power output obtained from LES.}
    \label{fig:powers}
\end{figure}

Finally, in \autoref{fig:powers}, the normalized turbine power outputs for the baseline RANS, LES, and perturbed models with data-free and data-driven approaches are given for Cases B and C. Fast wake recovery in the baseline model makes the effect of the upstream turbines' wake on the downstream counterpart seem insignificant, resulting in a higher power output compared to LES.
For UQ, in the data-free case $\delta$ = 1.0 value is used for the perturbation; however, in the data-driven cases, ML-predicted results with higher accuracy and less uncertainty show the ability of the data-driven method to predict other QoIs.

By analyzing the results obtained for Case B and Case C, one can conclude that the suggested framework performs well for various wind-farm layouts.
The generalizability of the trained ML model is equally impressive, and the features chosen as the model's input can estimate targets for unseen scenarios.
This denotes the independency of the framework and model from the turbine configurations in the wind farm.

\section{Conclusions} \label{sec:conc}
The leading goal of this study is to accurately quantify model-from uncertainties in predicting QoIs due to the Reynolds stress discrepancies for the RANS simulation of wind farms subjected to a neutrally stratified ABL condition. To this end, the realizable $k-\varepsilon$ turbulence model was used to solve the governing RANS equations, and the eigenvalue perturbation method was applied to the Reynolds stress anisotropy tensor for the wake-zone cells.
Based on a novel framework, in the first step, perturbation direction toward one of the turbulence limiting states was determined. The baseline RANS results were perturbed by the distance between the corresponding RANS and LES points on the barycentric map.
Perturbed model results with more accurate UQ confirmed the initial hypothesis about modifying the perturbation direction, utilizing a $priori$ determined perturbation value, and perturbing specified part of the domain in contrast to the data-free approach of the previous studies \cite{eidi2021model, hornshoj2021quantifying}. 

In the second step, based on the non-linearity of the eddy-viscosity and integrity basis tensors, physical explanations, and theoretical concepts, a set of 54 features was created for cells located in the wake zone. 
Then, by utilizing a two-step feature-selection technique, the necessary features were determined with the objectives of reducing the computational cost and preventing over-fitting. By comparing the performance of three different ML algorithms, the XGBoost algorithm as an advanced and emerging ensemble regression model was applied to predict the location of the LES anisotropy points for the cases in which the high-fidelity results are assumed to be unavailable.
We used data from a single wind-farm layout to train and test a model that could predict the results for two other unseen wind-farm cases: one with less turbine spacing, to incorporate increased turbulence intensity and enhanced wake effects; and the other one with the same streamwise turbine spacing but with non-aligned turbines to investigate partial wake effects in a more complex layout.

The trained model and the perturbation framework demonstrated generalizability and case-independency to a large extent when tested on the unseen cases. Normalized velocity deficit, turbulence intensity, and normalized turbines' power output were chosen as QoIs in this study. Results of the perturbed models covered the LES data when applied for the unseen cases in a similar manner as the training case.
This indicated that the proposed two-step feature selection was a key component and along with the selected ML algorithm prevents over- and under-fitting in estimating the target values. 
Further studies may seek to utilize emerging deep-learning techniques or alternative ML algorithms to quantify other forms of uncertainty, as well. 
In future works, various wind-farm or stand-alone turbine scenarios with diverse boundary conditions (e.g., hilly terrain and thermally stratified ABL) could be examined to extrapolate the ML model's capabilities in UQ estimations.

\section*{Acknowledgment}
N. Zehtabiyan-Rezaie and M. Abkar acknowledge the financial support from the Independent Research Fund Denmark (DFF) under the Grant No. 0217-00038B.

\section*{Conflict of interest}
The authors have no conflicts to disclose.

\section*{Data availability statement}
The data that support the findings of this study are available from the corresponding author upon reasonable request.

%\textcolor{blue}{
\appendix
%\counterwithin{figure}{section}
\section{Detection of the wake and free-stream zones} 
\label{app:A}
\begin{figure}%[htb]
    \centering
    \includegraphics[width=.7\linewidth]{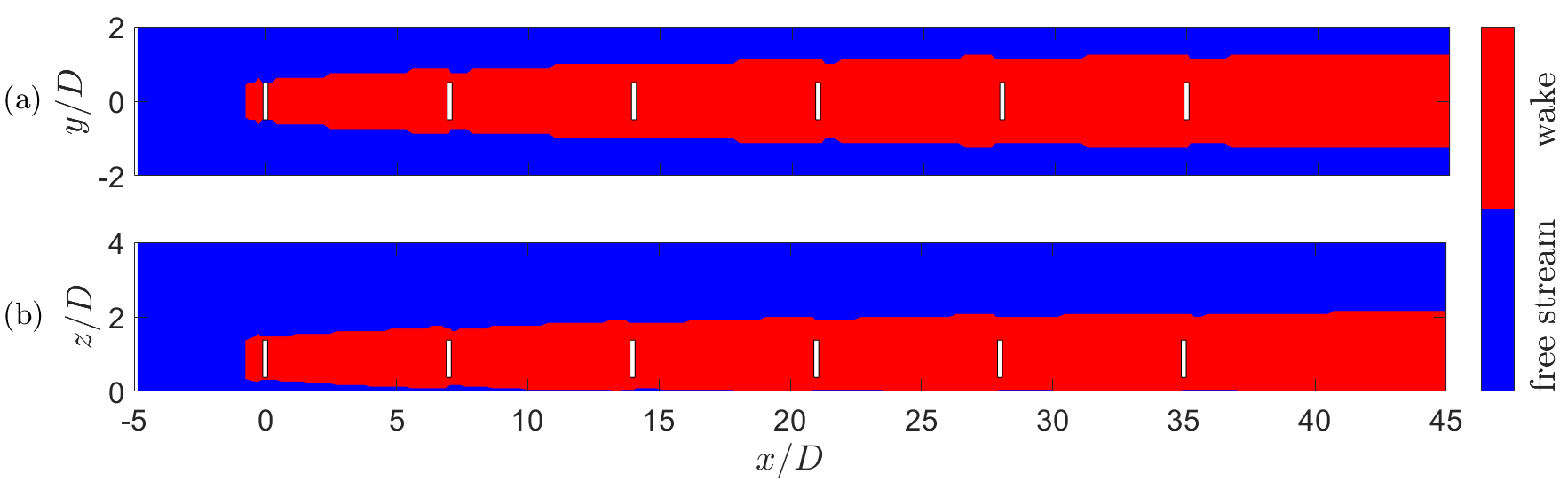}
    \caption{Detection of the wake and free stream zones at the hub height (a) and the  middle lateral plane of the domain in the spanwise direction (b) in Case A. The white rectangles indicate turbines.}
    \label{fig:wake}
\end{figure}

In this study, we evaluate perturbing either for all or a subset of the domain cells, as mentioned in the \autoref{sec:UQ md}.
The domain cells are divided into the wake and free-stream regions as

  \begin{equation}
         \label{eq:wake}
    \text{Cell location =}
    \begin{cases}
      \text{wake zone}, & \text{if}\ \Delta \bar{u}/\bar{u}_{h} > 0.02, \\
      \text{free stream zone}, & \text{otherwise.}
    \end{cases}
  \end{equation}
\autoref{fig:wake} indicates the wake and free-stream zones for Case A at the lateral plane crossing the middle of the domain in the spanwise direction at $z_{h}$.

\begin{figure}%[htb]
    \centering
    \includegraphics[width=0.7\linewidth]{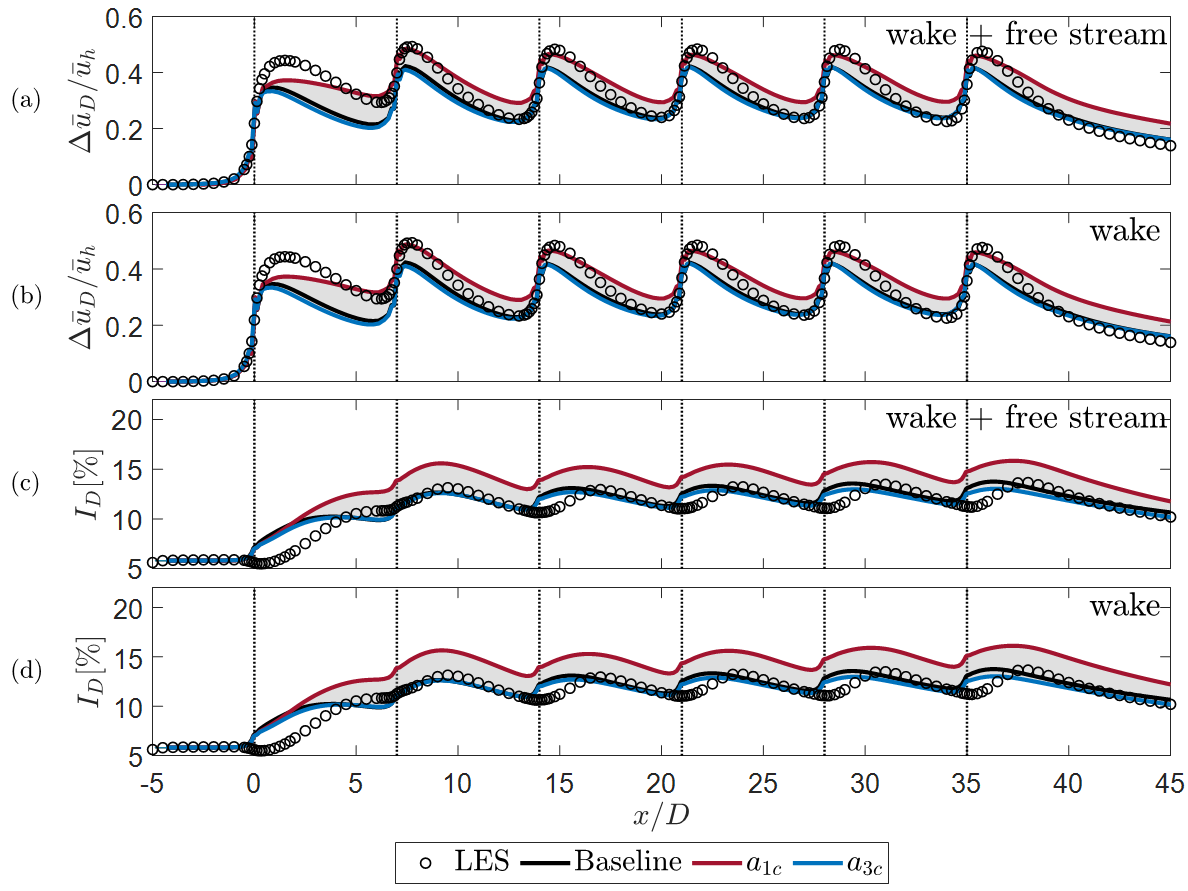}
    \caption{Effect of perturbation in the whole domain and wake zone on the rotor-averaged normalized velocity deficit  (a,b),  and turbulence intensity (c,d) in Case A.}
    \label{fig:wake_A_rotor}
\end{figure}

\autoref{fig:wake_A_rotor} depicts the normalized rotor-averaged velocity deficit and turbulence intensity in Case A for two cases: whole domain perturbation and wake-zone perturbation. It can be concluded that excluding the free-stream zone cells from the perturbation procedure has a negligible effect on the results.
Consequently, the use of the ML model for $\delta$ value prediction and excluding free-stream zone cells tend to decrease input data by more than 80\%, which has a significant impact on lowering computational cost.

\bibliographystyle{ieeetr} 
\bibliography{bibliography}

\end{document}